%% file: main.tex
\documentclass[journal,twoside,letterpaper]{IEEEtran}

\usepackage{amsmath,amsfonts}
\usepackage{algorithm}
\usepackage{algorithmic}
\usepackage{array}
\usepackage{textcomp}
\usepackage{stfloats}
\usepackage{url}
\usepackage{verbatim}
\usepackage{graphicx}
\def\BibTeX{{\rm B\kern-.05em{\sc i\kern-.025em b}\kern-.08em
    T\kern-.1667em\lower.7ex\hbox{E}\kern-.125emX}}
\usepackage{balance}

\usepackage{amssymb}
\usepackage{multirow}
\usepackage{svg}
\usepackage{booktabs}
\usepackage[font=footnotesize]{caption}
\usepackage{subcaption}
\usepackage{xcolor}

\usepackage[hidelinks]{hyperref}
\usepackage{orcidlink}

\usepackage[
    style=ieee,   
    ]{biblatex}
\usepackage{./resources/tikz/tikzit}
\input{./resources/tikz/default.tikzstyles}

\DeclareUnicodeCharacter{2212}{-}

\newcommand{\Description}[1]{}

\begin{document}
\title{Holistic Buoyancy-Driven Cloud Native Resource Management}

\author{Oliver~Larsson\orcidlink{0000-0003-0395-8313},
    Thijs~Metsch\orcidlink{0000-0003-3495-3646}, 
    Cristian~Klein\orcidlink{0000-0003-0106-3049}, 
    Erik~Elmroth\orcidlink{0000-0002-2633-6798},
\thanks{ 
    Funding for this project was provided 
    by the eSSENCE Programme under the
    Swedish Government’s Strategic Research Initiative.

    Experiments presented in this paper were carried out using 
    the Grid'5000 testbed, supported by a scientific 
    interest group hosted by Inria and including CNRS, 
    RENATER and several Universities as well as other 
    organizations (see \url{https://www.grid5000.fr}). 
}%
\thanks{Oliver Larsson, Cristian Klein, and Erik Elmroth are with the 
Department of Computing Science, Ume\aa{} University, SE-90187 Ume\aa, Sweden 
(e-mail: olars@cs.umu.se; cklein@cs.umu.se; elmroth@cs.umu.se).}%
\thanks{Thijs Metsch is unaffiliated, Germany 
(e-mail: tmetsch@engjoy.eu).}%
}

\maketitle

\begin{abstract}
  \input{content/00.abstract}

\end{abstract}


\input{./content/10.introduction.tex}
\input{./content/70.related-work.tex}
\input{./content/20.background.tex}

\input{./content/30.problem.tex}

\input{./content/40.design.tex}

\input{./content/50.implementation.tex}
\input{./content/55.experimental-setup.tex}
\input{./content/60.evaluation.tex}

\input{./content/80.conclusion.tex}

\section*{Acknowledgments}

Generative AI (including ChatGPT and Copilot) was used to improve the 
readability and language in the preparation of this work. The authors remain
solely responsible for its contents.

\printbibliography

\end{document}

%% file: resources/tikz/default.tikzstyles
\tikzstyle{Rectangle Medium}=[fill={rgb,255: red,42; green,71; blue,101}, draw=black, shape=rectangle, minimum width=2cm, text=white, minimum height=0.75cm, align=center]
\tikzstyle{Rectangle Large}=[fill={rgb,255: red,42; green,71; blue,101}, draw=black, shape=rectangle, text=white, minimum width=4cm, minimum height=2cm, align=center]
\tikzstyle{Red Text}=[fill=none, draw=none, shape=rectangle, tikzit draw={rgb,255: red,255; green,131; blue,131}, tikzit fill={rgb,255: red,255; green,131; blue,131}, text={{rgb,255: red,255; green,131; blue,131}}, align=center]
\tikzstyle{Small Circle}=[fill={rgb,255: red,42; green,71; blue,101}, draw=black, shape=circle, text=white]
\tikzstyle{Text}=[fill=none, draw=none, shape=rectangle, tikzit draw=black, align=center]
\tikzstyle{White Text}=[fill=none, draw=none, shape=rectangle, tikzit draw=black, align=center, text=white]
\tikzstyle{Rectangle Small}=[fill={rgb,255: red,42; green,71; blue,101}, draw=black, shape=rectangle, text=white, align=center, minimum height=0.5cm, minimum width=1.5cm]
\tikzstyle{Rectangle MedLarge}=[fill={rgb,255: red,42; green,71; blue,101}, draw=black, shape=rectangle, text=white, align=center, minimum height=1.33cm, minimum width=2.5cm]
\tikzstyle{Small Outline Circle}=[fill=white, draw=black, shape=circle, inner sep=0pt]

\tikzstyle{Arrow}=[draw=black, ->, line width=.025cm]
\tikzstyle{Edge}=[-, draw=black, line width=0.025cm, fill=none]
\tikzstyle{Red Dashed}=[-, draw={rgb,255: red,255; green,131; blue,131}, dashed, line width=0.025cm]
\tikzstyle{Dashed}=[-, draw={rgb,255: red,136; green,136; blue,136}, dashed, line width=0.025cm]
\tikzstyle{Freebox}=[-, draw=black, fill={rgb,255: red,42; green,71; blue,101}]
\tikzstyle{Freebox Green}=[-, draw=black, fill={rgb,255: red,115; green,167; blue,144}]
\tikzstyle{Freebox Yellow}=[-, draw=black, fill={rgb,255: red,215; green,177; blue,124}]
\tikzstyle{Freebox Red}=[-, draw=black, fill={rgb,255: red,255; green,131; blue,131}]
\tikzstyle{Thick Dashed}=[-, line width=0.025cm, draw=black, dashed]
\tikzstyle{Black Lines}=[-, fill=none, pattern=north east lines, pattern color={{rgb,255: red,42; green,71; blue,101}}, draw=black, tikzit fill={rgb,255: red,191; green,191; blue,191}, tikzit draw={rgb,255: red,241; green,239; blue,228}, line width=0.025cm]
\tikzstyle{Blue Background}=[-, fill={{rgb,255: red,42; green,71; blue,101}}, fill opacity=0.3, draw=none]
\tikzstyle{Green Background}=[-, fill={{rgb,255: red,115; green,167; blue,144}}, fill opacity=0.45, draw=none]
\tikzstyle{Black Lines Dashed}=[-, fill=none, pattern=north east lines, pattern color={{rgb,255: red,42; green,71; blue,101}}, draw=black, tikzit fill={rgb,255: red,191; green,191; blue,191}, tikzit draw={rgb,255: red,241; green,239; blue,228}, line width=0.025cm, dashed]
\tikzstyle{Blue Dots}=[-, fill=none, pattern=horizontal lines, pattern color={{rgb,255: red,215; green,177; blue,124}}, tikzit fill={rgb,255: red,42; green,71; blue,101}, line width=0.025cm]
\tikzstyle{Red Bricks}=[-, fill=none, pattern=crosshatch, line width=0.025cm, tikzit fill={rgb,255: red,255; green,131; blue,131}, pattern color={{rgb,255: red,115; green,167; blue,144}}]
\tikzstyle{Black Cross}=[-, fill=none, pattern=crosshatch dots, line width=0.025cm, tikzit fill={rgb,255: red,255; green,131; blue,131}, pattern color={{rgb,255: red,42; green,71; blue,101}}]
\tikzstyle{Red Bricks Dashed}=[-, fill=none, pattern=crosshatch, line width=0.025cm, tikzit fill={rgb,255: red,255; green,131; blue,131}, pattern color={{rgb,255: red,115; green,167; blue,144}}, dashed]
\tikzstyle{Dashed Arrow}=[->, line width=0.025cm, draw=black, dashed]
\tikzstyle{Thin Dashed Arrow}=[->, line width=0.025cm, dashed, draw={rgb,255: red,136; green,136; blue,136}]
\tikzstyle{Blue EC F}=[-, draw={rgb,255: red,42; green,71; blue,101}, fill={rgb,255: red,42; green,71; blue,101}, fill opacity=0.33, line width=0.25mm]
\tikzstyle{Green EC F}=[-, draw={rgb,255: red,115; green,167; blue,144}, fill={rgb,255: red,115; green,167; blue,144}, fill opacity=0.33, line width=0.25mm]
\tikzstyle{Yellow EC F}=[-, draw={rgb,255: red,215; green,177; blue,124}, fill={rgb,255: red,215; green,177; blue,124}, fill opacity=0.33, line width=0.25mm]
\tikzstyle{Pink EC F}=[-, draw={rgb,255: red,234; green,186; blue,185}, fill={rgb,255: red,234; green,186; blue,185}, fill opacity=0.33, line width=0.25mm]

%% file: content/00.abstract.tex
Efficient resource management is a challenge in multi-tenant cloud environments 
as workloads compete for shared resources such as last-level cache and memory bandwidth, whose congestion can 
change dynamically as workload composition and resource demands evolve. 
Existing resource management approaches often rely on static resource 
requirements or prior workload profiling, limiting their 
ability to respond to changing resource contention as required in the cloud. 
In this paper, we leverage workload buoyancy, a runtime framework for capturing 
workload sensitivity to individual resources and application-level 
performance, to enable adaptive scheduling and resource management. 
We propose a holistic approach that incorporates buoyancy into 
workload scheduling, node-local resource allocation, 
and workload rescheduling throughout the workload lifecycle. 
We implement our approach as a Kubernetes-native prototype and evaluate 
it on a multi-node cluster using diverse cloud workloads under varying 
levels of resource contention. In a high-congestion scenario, 
our approach increases the percentage of workloads satisfying their SLOs 
from 7.74\% to 87.8\% on average compared with the Kubernetes baseline, 
while reducing resource congestion, the severity of SLO violations, 
and cluster fragmentation using limited rescheduling.

%% file: content/10.introduction.tex
\section{Introduction}

The rapid growth of AI and other compute-intensive workloads is increasing 
the demand for computing infrastructure, making efficient utilization of cloud 
compute clusters increasingly important. 
Cloud providers must therefore maximize the utilization of their available hardware without violating 
the performance expectations of their tenants. 
This is challenging because the performance of a workload depends not only on the resources 
allocated to it, but also on contention for shared system resources such as last-level 
cache (LLC), and other interactions between co-located 
workloads~\cite{abediConductingRepeatableExperiments2017,freischuetzTUNATuningUnstable2025}. 

This creates a fundamental conflict between the goals of involved stakeholders.
\emph{Application owners}, who primarily care about meeting the performance requirements of their workloads,
generally allocate more resources than they consume in order to 
provide sufficient performance headroom. 
On the other hand, \emph{resource owners} seek 
to consolidate workloads and maximize the utilization of their hardware.
To do so, they commonly overcommit their clusters and 
sell more capacity than is physically available~\cite{bashir_take_2021}. 
While overcommitment can improve resource utilization, it also increases the 
likelihood of contention as more workloads compete for the same shared resources. 
Consequently, application owners often compensate for this unpredictable 
performance by allocating even more resources, beyond their actual requirements, 
reducing the effective utilization of the cluster and creating a vicious cycle 
of overprovisioning and overcommitment that originates from the conflicting goals
of the stakeholders.

Previous work has demonstrated that resource allocation and isolation techniques 
can mitigate the effects of such contention~\cite{chen_parties_2019,patel_clite_2020,chen_olpart_2023}. 
However, such techniques generally do not consider contention over individual system resources
or the conflicting goals of stakeholders.
Furthermore, cloud environments are inherently dynamic: workloads are continuously added and removed, 
and their resource demands change over time. Consequently, 
resource management decisions that are effective at one point in time may become 
suboptimal as workload behavior and contention patterns change. 
Failing to account for these dynamics can result in cluster 
fragmentation and resource stranding, reducing cluster utilization and potentially 
requiring rescheduling to recover stranded capacity~\cite{larsson_impact_2023}.

Effective resource management therefore requires the resource owner to understand 
not only how much of each resource a workload has been allocated, 
but also how workloads interact with, and are affected by, contention for any given 
shared system resource. The \emph{workload buoyancy} monitoring framework provides such insights
by defining metrics that quantify the resource contention experienced 
by workloads~\cite{larssonWorkloadBuoyancyKeeping2026}. These metrics provide a 
basis for identifying workload sensitivity to resource contention and 
for understanding how contention evolves as workloads compete for cluster 
resources. We argue that this information can be used to improve cluster 
resource management by incorporating observed workload congestion into 
scheduling, rescheduling, and resource allocation decisions.

In this paper, we investigate how workload buoyancy can be leveraged to improve 
the utilization and performance of cloud compute clusters. We propose a 
holistic three-component resource management approach that uses buoyancy 
metrics to inform \emph{(i)} workload scheduling, \emph{(ii)} rescheduling 
of existing workloads, and \emph{(iii)} dynamic node-local allocation of resources
such as CPU and LLC.
Together, these mechanisms allow the resource owner to adapt resource placement 
and allocation to observed contention while respecting the resource requirements 
and performance constraints of tenant workloads. Specifically, we make the following contributions:
\begin{itemize}
\item We propose a resource management approach that leverages workload buoyancy 
    metrics to inform scheduling, rescheduling, and dynamic node-local resource allocation in cloud compute clusters.
\item We demonstrate that incorporating workload congestion into resource management 
    decisions can substantially improve workload service-level objective (SLO) 
    compliance while reducing resource contention and cluster fragmentation.
\item We implement a proof-of-concept prototype of the proposed approach and evaluate it in a Kubernetes-based cloud environment.
\end{itemize}

Our evaluation demonstrates that the proposed approach substantially improves 
SLO compliance under high resource congestion. Compared with a Kubernetes baseline, 
the approach increases the average percentage of workloads operating within 
their SLOs from $7.74\%$ to $87.8\%$, while simultaneously reducing the magnitude 
of SLO violations when they occur. 
These results demonstrate the potential of workload-level congestion information 
as a signal for resource management in dynamic, multi-tenant cloud environments.

%% file: content/70.related-work.tex
\section{Related Work}

Resource management in the cloud has been extensively studied across 
workload scheduling, resource allocation, performance 
interference, and runtime adaptation. We focus on two areas most 
closely related to our approach: application-aware scheduling 
and rescheduling, and runtime resource management and isolation.

\subsubsection{Application-Aware Scheduling and Rescheduling}

Application-aware scheduling incorporates information about 
workload behavior or application-level performance into 
placement and resource-management decisions. 
Quality of service (QoS) and quality of experience (QoE) aware 
approaches use application-level 
performance insights to guide workload placement and 
allocation~\cite{da_silva_qos-driven_2020,carvalho_qoe_2021}, 
while other approaches use predicted workload behavior to make 
proactive scheduling decisions~\cite{giannakopoulos_performance-aware_2025}. 
Runtime resource utilization has similarly been used to improve cluster-level 
scheduling decisions~\cite{zhang_zeus_2021,aindalaAVOAEnergyResource2025}. 
However, resource utilization does not necessarily indicate the performance 
impact of contention on a workload. Workload buoyancy addresses this limitation 
by combining application-level performance with resource sensitivity, and 
has previously been shown to improve workload placement~\cite{larsson_beyond_2026}. 
In this work, we extend this approach beyond initial placement to also 
inform runtime resource management and rescheduling.

\subsubsection{Runtime Resource Management and Isolation}

Runtime resource management approaches adapt resource allocations or 
enforce isolation based on workload requirements and system state. 
Many approaches rely on workload profiling and resource 
models~\cite{park_copart_2019,zhao_rhythm_2020}, while optimization-based 
methods can identify efficient resource allocations through iterative 
exploration~\cite{chen_parties_2019,roySATORIEfficientFair2021}. Other approaches 
instead use runtime resource utilization~\cite{chen_olpart_2023} or 
application-level performance metrics~\cite{larsson_esther_2025} to 
dynamically adjust allocations. Hardware-level isolation mechanisms 
can further mitigate interference by controlling access to shared 
resources~\cite{larssonHardwareLevelQoSEnforcement2025}. These approaches 
demonstrate the benefits of runtime adaptation, but typically 
focus on resource allocation or isolation within a node and do not 
coordinate such decisions with cluster-level workload placement.

\subsubsection{Summary}

In contrast, our approach uses workload buoyancy as a 
common runtime signal for both workload performance and 
resource sensitivity. We use this information to coordinate initial 
scheduling, node-local resource allocation, and rescheduling, 
allowing resource-management decisions to adapt to changing workload 
composition and resource contention without requiring prior workload profiling.

%% file: content/20.background.tex
\section{Background}

Resource management in a multi-tenant cloud environment requires the 
resource owner to allocate resources and place workloads while accounting for 
both resource capacity and workload performance. Resource requests 
and utilization provide information about how resources are allocated and 
consumed, but do not necessarily indicate how resource contention 
affects workload performance. This distinction is particularly important for 
shared system resources such as the LLC, where workloads can interfere with one another 
despite having sufficient allocated capacity.

\subsection{Resource Management and Workload Observability}

Application owners generally provision workloads with sufficient resource 
headroom to tolerate variations in demand and resource contention, and in some
cases, to ``draw in'' enough system resources such as LLC or memory bandwidth as
they are generally not explicitly allocatable.
Resource owners, in contrast, seek to consolidate workloads to 
maximize hardware utilization. These objectives can conflict when increasing 
utilization through overcommitment causes additional contention 
and consequently degrades workload performance.

The resource owner is ultimately responsible for deciding how workloads 
are placed and how resources are allocated. However, resource owners 
typically have limited visibility into the performance characteristics and 
resource sensitivity of tenant workloads. Infrastructure-level measurements 
such as resource requests and utilization describe resource consumption, 
but provide limited information about whether a workload is sensitive to 
contention or whether contention is causing it to violate its SLO.

Prior work has demonstrated that application-level performance information and 
resource characteristics can improve both node-local resource 
management~\cite{zhao_rhythm_2020,larsson_esther_2025} and cluster-level 
scheduling~\cite{da_silva_qos-driven_2020,carvalho_qoe_2021,giannakopoulos_performance-aware_2025}. 
Similarly, identifying the resources that constrain individual workloads can improve 
resource allocation and placement decisions~\cite{zhang_zeus_2021,chen_olpart_2023,larsson_beyond_2026}.

However, obtaining and maintaining this information is challenging in dynamic multi-tenant environments.

\subsection{Challenges of Runtime Resource Characterization}

Detailed workload characterization often requires application profiling to model 
resource requirements and interactions between different 
resources~\cite{byrne_survey_2018,park_copart_2019,zhao_rhythm_2020}. However, 
such profiles may become inaccurate as workloads change their behavior or 
are deployed alongside different workloads.

Moreover, a workload's performance depends not only on its own resource 
requirements but also on the workloads with which it is colocated. A resource 
profile obtained in isolation may therefore provide an incomplete picture of the 
congestion experienced in a shared environment. As the workload composition of 
a node changes, the same workload may experience different levels of 
resource contention even when its own resource demand remains unchanged.

Together, these characteristics make continuous workload characterization 
impractical in general-purpose cloud environments. Effective resource management 
instead requires runtime observability that captures the relationship between 
workload performance, resource sensitivity, and resource contention without 
relying on extensive prior profiling.

\subsection{Workload Buoyancy}

The \emph{workload buoyancy} framework addresses these observability 
challenges by providing an abstraction that can be leveraged by both application 
owners and resource owners. It is a 
runtime representation of workload sensitivity to 
resource contention~\cite{larssonWorkloadBuoyancyKeeping2026}. Rather than relying 
solely on resource utilization or requiring prior application profiling, 
workload buoyancy combines system-level resource observations with 
application-level service-level indicators (SLI) to characterize the 
performance headroom available to workloads.

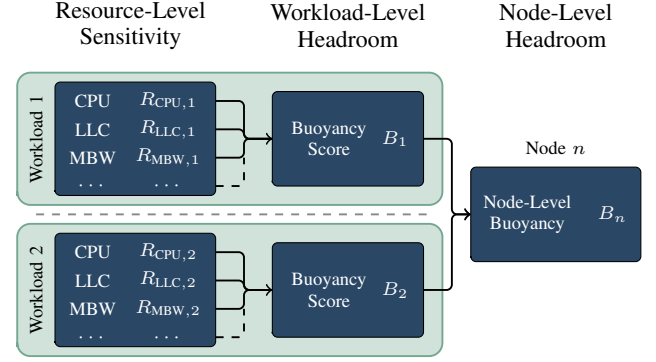
\begin{figure}[tbp]
\centering
\input{buoyancy.tikz}
\caption{Resource scores, buoyancy scores, and node-level buoyancy 
    scores at different context levels.}
\label{fig:buoyancy-scores}
\end{figure}

Buoyancy models resource sensitivity at multiple levels of abstraction. 
At the hardware-resource level, \emph{resource scores} are computed for 
resources of interest, such as CPU, LLC, memory bandwidth, and 
I/O. A resource score is a value in the range $[0,1]$ representing the 
sensitivity of an application to a particular resource, where $0$ indicates 
low sensitivity and $1$ indicates high sensitivity.
At the application level, a \emph{buoyancy score} is derived from the application's 
resource scores and application-level SLIs. The buoyancy score lies in 
the range $(-\infty,1]$ and represents the performance 
headroom available to the application. Intuitively, a workload with high buoyancy can 
tolerate additional resource pressure, whereas a workload with low buoyancy 
is more susceptible to performance degradation. A negative buoyancy score 
indicates that the workload is currently violating its SLO.
Finally, a \emph{node-level buoyancy score} aggregates the buoyancy of workloads 
executing on a node. This provides a workload-centric indication of the 
performance headroom of the node and can be used to reason about the 
effects of adding, removing, or reallocating workloads as the node's workload composition changes.
The three levels at which buoyancy is computed are illustrated in \figurename~\ref{fig:buoyancy-scores}.
For a more detailed description of the buoyancy framework, we refer the reader to
the original work by Larsson et al.~\cite{larssonWorkloadBuoyancyKeeping2026}.

The resulting metrics provide complementary information for resource management. 
Resource scores characterize the resources to which individual workloads are sensitive, 
workload buoyancy captures the performance headroom of individual workloads, 
and node-level buoyancy provides an aggregate view of the performance headroom of 
colocated workloads. Together, these signals allow resource-management 
decisions to account for both workload characteristics and the current workload composition of a node.

In this work, we use these signals as inputs to three resource-management 
mechanisms: workload scheduling, runtime rescheduling, and dynamic node-local resource allocation.
While buoyancy metrics have been shown to quantify the performance headroom 
of workloads and nodes with respect to individual resources~\cite{larssonWorkloadBuoyancyKeeping2026}, 
their use in improving holistic, cluster-wide resource management while respecting the goals
of involved stakeholders has not been previously explored.

%% file: resources/tikz/buoyancy.tikz
\begin{tikzpicture}
	\begin{pgfonlayer}{nodelayer}
		\node [style=none] (0) at (-5.25, 2) {};
		\node [style=none] (1) at (-5.25, -1) {};
		\node [style=none] (2) at (-1, -1) {};
		\node [style=none] (3) at (-1, 2) {};
		\node [style=none] (48) at (-5.75, -1.5) {};
		\node [style=none] (49) at (4.75, -1.5) {};
		\node [style=none] (50) at (-1, -2.5) {};
		\node [style=none] (51) at (-0.25, -2.5) {};
		\node [style=none] (52) at (-0.25, -3.5) {};
		\node [style=none] (53) at (0.5, -3.5) {};
		\node [style=none] (54) at (-1, -3.25) {};
		\node [style=none] (55) at (-0.25, -3.25) {};
		\node [style=none] (56) at (-1, -4) {};
		\node [style=none] (57) at (-0.25, -4) {};
		\node [style=none] (58) at (-1, -4.75) {};
		\node [style=none] (59) at (-0.25, -4.75) {};
		\node [style=none] (60) at (-1, 1.5) {};
		\node [style=none] (61) at (-0.25, 1.5) {};
		\node [style=none] (62) at (-0.25, 0.5) {};
		\node [style=none] (63) at (0.5, 0.5) {};
		\node [style=none] (64) at (-1, 0.75) {};
		\node [style=none] (65) at (-0.25, 0.75) {};
		\node [style=none] (66) at (-1, 0) {};
		\node [style=none] (67) at (-0.25, 0) {};
		\node [style=none] (68) at (-1, -0.75) {};
		\node [style=none] (69) at (-0.25, -0.75) {};
		\node [style=none] (70) at (4.5, 0.5) {};
		\node [style=none] (71) at (5.25, 0.5) {};
		\node [style=none] (72) at (5.25, -1.5) {};
		\node [style=none] (73) at (5.75, -1.5) {};
		\node [style=none] (74) at (4.5, -3.5) {};
		\node [style=none] (75) at (5.25, -3.5) {};
		\node [style=none] (76) at (5.25, -1.5) {};
		\node [style=none] (77) at (5.75, -1.5) {};
		\node [style=none] (80) at (0.5, 1.75) {};
		\node [style=none] (81) at (0.5, -0.75) {};
		\node [style=none] (82) at (4.5, -0.75) {};
		\node [style=none] (83) at (4.5, 1.75) {};
		\node [style=none] (84) at (-5.25, -2) {};
		\node [style=none] (85) at (-5.25, -5) {};
		\node [style=none] (86) at (-1, -5) {};
		\node [style=none] (87) at (-1, -2) {};
		\node [style=none] (88) at (0.5, -2.25) {};
		\node [style=none] (89) at (0.5, -4.75) {};
		\node [style=none] (90) at (4.5, -4.75) {};
		\node [style=none] (91) at (4.5, -2.25) {};
		\node [style=none] (92) at (5.75, -0.25) {};
		\node [style=none] (93) at (5.75, -2.75) {};
		\node [style=none] (94) at (10.25, -2.75) {};
		\node [style=none] (95) at (10.25, -0.25) {};
		\node [style=none] (96) at (-6.25, 2.25) {};
		\node [style=none] (97) at (-6.25, -1.25) {};
		\node [style=none] (98) at (5, -1.25) {};
		\node [style=none] (99) at (5, 2.25) {};
		\node [style=none] (100) at (-6.25, -1.75) {};
		\node [style=none] (101) at (-6.25, -5.25) {};
		\node [style=none] (102) at (5, -5.25) {};
		\node [style=none] (103) at (5, -1.75) {};
		\node [style=Text, font={\small}] (12) at (-3.25, 3.5) {Resource-Level\\Sensitivity};
		\node [style=Text, font={\small}] (13) at (2.5, 3.5) {Workload-Level\\Headroom};
		\node [style=Text, font={\small}] (14) at (8, 3.5) {Node-Level\\Headroom};

		\node [style=Text, rotate=90] (23) at (-5.75, 0.5) {\scriptsize Workload 1};
		\node [style=White Text] (15) at (-4.25, 1.5) {\scriptsize CPU};
		\node [style=White Text] (16) at (-2.25, 1.5) {\scriptsize $R_{\text{CPU},1}$};
		\node [style=White Text] (17) at (-4.25, 0.75) {\scriptsize LLC};
		\node [style=White Text] (18) at (-2.25, 0.75) {\scriptsize $R_{\text{LLC},1}$};
		\node [style=White Text] (19) at (-4.25, 0) {\scriptsize MBW};
		\node [style=White Text] (20) at (-2.25, 0) {\scriptsize $R_{\text{MBW},1}$};
		\node [style=White Text] (37) at (-4.25, -0.75) {\scriptsize $\cdots$};
		\node [style=White Text] (38) at (-2.25, -0.75) {\scriptsize $\cdots$};
		\node [style=White Text, font={\scriptsize}] (40) at (2, 0.5) {Buoyancy\\Score};
		\node [style=White Text, font={\scriptsize}] (41) at (3.75, 0.5) {$B_1$};
		\node [style=Text] (108) at (8, 0.25) {\scriptsize Node $n$};
		\node [style=White Text, font={\scriptsize}] (78) at (7.25, -1.5) {Node-Level\\Buoyancy};
		\node [style=White Text, font={\scriptsize}] (79) at (9.5, -1.5) {$B_n$};

		\node [style=Text, rotate=90] (36) at (-5.75, -3.5) {\scriptsize Workload 2};
		\node [style=White Text] (28) at (-4.25, -2.5) {\scriptsize CPU};
		\node [style=White Text] (29) at (-2.25, -2.5) {\scriptsize $R_{\text{CPU},2}$};
		\node [style=White Text] (30) at (-4.25, -3.25) {\scriptsize LLC};
		\node [style=White Text] (31) at (-2.25, -3.25) {\scriptsize $R_{\text{LLC},2}$};
		\node [style=White Text] (32) at (-4.25, -4) {\scriptsize MBW};
		\node [style=White Text] (33) at (-2.25, -4) {\scriptsize $R_{\text{MBW},2}$};
		\node [style=White Text] (34) at (-4.25, -4.75) {\scriptsize $\cdots$};
		\node [style=White Text] (39) at (-2.25, -4.75) {\scriptsize $\cdots$};
		\node [style=White Text, font={\scriptsize}] (46) at (2, -3.5) {Buoyancy\\Score};
		\node [style=White Text, font={\scriptsize}] (47) at (3.75, -3.5) {$B_2$};
	\end{pgfonlayer}
	\begin{pgfonlayer}{edgelayer}
		\draw [style=Green EC F, rounded corners=5pt] (97.center)
			 to (98.center)
			 to (99.center)
			 to (96.center)
			 to cycle;
		\draw [style=Green EC F, rounded corners=5pt] (101.center)
			 to (102.center)
			 to (103.center)
			 to (100.center)
			 to cycle;
		\draw [style=Freebox, rounded corners=2pt] (1.center)
			 to (2.center)
			 to (3.center)
			 to (0.center)
			 to cycle;
		\draw [style=Dashed] (48.center) to (49.center);
		\draw [style=Edge, rounded corners=2pt] (54.center)
			 to (55.center)
			 to (52.center)
			 to (53.center);
		\draw [style=Edge, rounded corners=2pt] (56.center)
			 to (57.center)
			 to (52.center)
			 to (53.center);
		\draw [style=Thick Dashed, rounded corners=2pt] (58.center)
			 to (59.center)
			 to (57.center);
		\draw [style=Edge, rounded corners=2pt] (64.center)
			 to (65.center)
			 to (62.center)
			 to (63.center);
		\draw [style=Edge, rounded corners=2pt] (66.center)
			 to (67.center)
			 to (62.center)
			 to (63.center);
		\draw [style=Thick Dashed, rounded corners=2pt] (68.center)
			 to (69.center)
			 to (67.center);
		\draw [style=Freebox, rounded corners=2pt] (81.center)
			 to (82.center)
			 to (83.center)
			 to (80.center)
			 to cycle;
		\draw [style=Freebox, rounded corners=2pt] (85.center)
			 to (86.center)
			 to (87.center)
			 to (84.center)
			 to cycle;
		\draw [style=Freebox, rounded corners=2pt] (89.center)
			 to (90.center)
			 to (91.center)
			 to (88.center)
			 to cycle;
		\draw [style=Freebox, rounded corners=2pt] (93.center)
			 to (94.center)
			 to (95.center)
			 to (92.center)
			 to cycle;
		\draw [style=Arrow, rounded corners=2pt] (50.center)
			 to (51.center)
			 to (55.center)
			 to (52.center)
			 to (53.center);
		\draw [style=Arrow, rounded corners=2pt] (60.center)
			 to (61.center)
			 to (65.center)
			 to (62.center)
			 to (63.center);
		\draw [style=Arrow, rounded corners=2pt] (70.center)
			 to (71.center)
			 to (72.center)
			 to (73.center);
		\draw [style=Arrow, rounded corners=2pt] (74.center)
			 to (75.center)
			 to (76.center)
			 to (77.center);
	\end{pgfonlayer}
\end{tikzpicture}

%% file: content/30.problem.tex
\section{Problem Statement}

Resource owners must continuously make resource-management 
decisions despite incomplete knowledge of workload behavior and changing resource 
contention. Resource requests and utilization provide information about resource 
allocation and consumption, but do not directly indicate the sensitivity of 
workloads to contention or the performance impact of colocating workloads. 
Workload buoyancy provides runtime information that can bridge this gap 
by relating workload performance to sensitivity across individual system resources
in an extensible and future-proof manner.

We therefore consider the problem of incorporating workload buoyancy into 
resource-management decisions in dynamic, multi-tenant cloud environments. 
The resource-management approach should use runtime buoyancy insights to 
account for both the current and expected effects of resource contention 
while adapting to changes in workload composition and resource demand.

More specifically, the approach should provide three complementary capabilities:

\begin{enumerate}
\item \textbf{Contention-aware scheduling:} When a new workload is deployed, 
    use the buoyancy of existing workloads and the resource sensitivity 
    of the new workload to select a placement that minimizes the expected impact of resource contention.

\item \textbf{Runtime rescheduling:} Detect changes in workload congestion and identify 
    placements that cause harmful contention, allowing workloads to be rescheduled when 
    their current placement becomes detrimental to performance or cluster efficiency.

\item \textbf{Dynamic resource allocation:} Adapt node-local resource allocations 
    according to the resource sensitivity and performance requirements of colocated 
    workloads, reducing contention while making use of otherwise underutilized resources.

\end{enumerate}

These mechanisms should operate without requiring prior profiling of individual 
workloads and should adapt as workloads enter, leave, or change their 
resource behavior. The overall objective is to improve workload SLO 
compliance and cluster resource utilization while reducing harmful resource 
contention and resource fragmentation.

%% file: content/40.design.tex
\section{Solution Design} \label{sec:design}

Our solution consists of three logically independent components that use workload 
buoyancy to make resource management decisions at different stages of 
the workload lifecycle: \emph{scheduling}, \emph{resource management}, 
and \emph{rescheduling}. The components are designed to operate independently 
of the underlying cloud platform or orchestrator. Their responsibilities and 
decision-making policies are defined in terms of workload- and node-level state, 
allowing the policies to be implemented using platform-specific mechanisms.

\begin{figure}[tbp]
\centering
\input{approach.tikz}
\caption{Overview of the proposed buoyancy-aware resource management approach. 
Scheduling uses node-level buoyancy to guide initial workload placement. 
Resource management continuously adjusts resource allocations, while rescheduling 
changes workload placement when local resource management cannot resolve resource pressure.}
\label{fig:approach}
\end{figure}
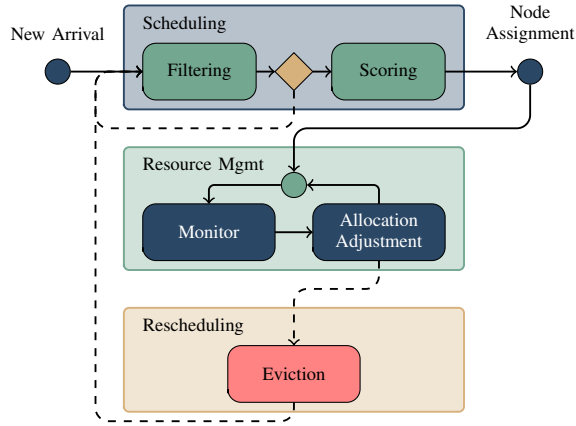

\figurename~\ref{fig:approach} provides an overview of the proposed approach. 
When a new workload is submitted, the \emph{scheduling component} uses 
node-level buoyancy to guide its placement. Once workloads are 
running, the \emph{resource management component} continuously monitors the 
workloads on each node and adjusts their resource allocations based on 
their observed buoyancy and resource scores. If resource pressure cannot be 
resolved through local resource allocation, the \emph{rescheduling component} 
selects a workload to move away from the affected node. The workload is then 
returned to the scheduling process, where it can be placed on another 
node based on the current state of the cluster.

The three components therefore form a feedback loop. Scheduling determines 
the initial placement of workloads, resource management adapts resource allocations 
as workload behavior changes, and rescheduling changes workload placement when 
local resource management is insufficient. This enables resource management 
decisions to adapt continuously to changes in workload composition and resource demand.

The design assumes that workload and node buoyancy metrics, as defined by the 
workload buoyancy framework~\cite{larssonWorkloadBuoyancyKeeping2026}, are available 
to the resource management components. We denote the buoyancy score of 
workload $w$ as $B_w$ and the buoyancy score of node $n$ as $B_n$. Similarly, $R_{r,w}$ 
denotes the resource score of workload $w$ for resource $r$, while $R_{r,n}$ 
denotes the resource score of node $n$ for resource $r$.

The remainder of this section describes the decision-making policies of the 
three components. Platform-specific mechanisms used to realize 
these policies are described in Section~\ref{sec:implementation}.

\subsection{Scheduling Component}\label{sec:scheduling-design}

The scheduling component guides the orchestrator in selecting a node for 
newly submitted workloads. Its objective is to avoid placing workloads on 
nodes that are already experiencing resource pressure and to prefer nodes with 
greater performance headroom to accommodate additional workload demand.

The scheduling component follows a two-stage process consisting of 
\emph{filtering} and \emph{scoring}. During filtering, candidate nodes are 
evaluated to determine which nodes are feasible for scheduling. During scoring, 
the feasible nodes are ranked according to their suitability. The orchestrator 
then selects the highest-ranked node for workload placement.

An important design choice is that the scheduling component uses the node-level 
buoyancy of candidate nodes rather than requiring a detailed prediction of the 
interaction between a new workload and the workloads already executing on each node. 
Node-level buoyancy provides an aggregate indication of the performance headroom 
available on a node and therefore allows placement decisions to be made without 
prior profiling of the workload being scheduled.

\subsubsection{Filtering}

We use node-level buoyancy as an indicator of the ability of a node to 
accommodate additional workload demand. Previous work has shown that node-level 
buoyancy can be used to produce less congested workload distributions~\cite{larsson_beyond_2026}.

Let $N$ denote the set of nodes in the cluster. We define a filtering function
$F$ as
\begin{equation}
F(n) =
\begin{cases}
\textrm{true}, & B_n \geq 0.1,\\
\textrm{false}, & \text{otherwise}.
\end{cases}
\end{equation}
A node is therefore considered feasible if its current buoyancy score is at 
least $0.1$. This threshold was experimentally selected to exclude nodes 
whose existing workloads have limited performance headroom that would make them
breach their performance targets if additional workloads were placed on them.
Such nodes are thus considered unsuitable for accommodating additional resource demand.

\subsubsection{Scoring} \label{sec:scoring}

After filtering, feasible nodes are ranked according to their buoyancy. 
We define a scoring function $S$ that assigns higher scores to nodes with greater buoyancy. 
This encourages the scheduler to place new workloads on nodes with greater performance headroom,
spreading workloads across available nodes according to their ability to accommodate 
additional resource demand. The exact range of the scoring function depends on the requirements of the underlying 
orchestrator. In general, the buoyancy score is mapped to the scoring range 
supported by the orchestrator. In our implementation, we map the buoyancy scores to
integer scores in the range $[0,10]$; the corresponding implementation is described in Section~\ref{sec:implementation}.
For the purposes of the design, we define the score as
\begin{equation}
S(n) =
\begin{cases}
\lceil 10B_n \rceil, & B_n > 0,\\
0, & \text{otherwise}.
\end{cases}
\end{equation}
Because $B_n \in (-\infty,1]$, this formulation produces scores in the range $[0,10]$ 
while preserving the ordering of nodes with positive buoyancy. 
The exact scaling factor is an implementation parameter and does not 
affect the underlying scheduling policy.

\subsection{Resource Management Component} \label{sec:resource-management}

The resource management component continuously adapts the resource allocations of 
workloads executing on each node. It operates as a feedback controller that observes 
workload performance headroom and resource sensitivity and uses this information to redistribute resources between workloads.

The component attempts to achieve two complementary objectives. First, it should 
provide additional resources to workloads experiencing performance pressure. 
Second, it should reclaim resources from workloads with sufficient performance 
headroom so that these resources can be used by other workloads. If the controller 
cannot resolve resource pressure through local reallocation, it notifies the rescheduling 
component which in turn takes rescheduling action.

\begin{figure}[tbp]
\centering
\input{resource-manager.tikz}
\caption{High-level operation of the node-local resource management controller. 
The controller observes workload state, classifies workloads according to buoyancy, 
reclaims resources where possible, and attempts to increase resources for 
workloads experiencing performance pressure. If local resource reallocation cannot 
resolve the pressure, rescheduling is triggered.}
\label{fig:node-level}
\end{figure}
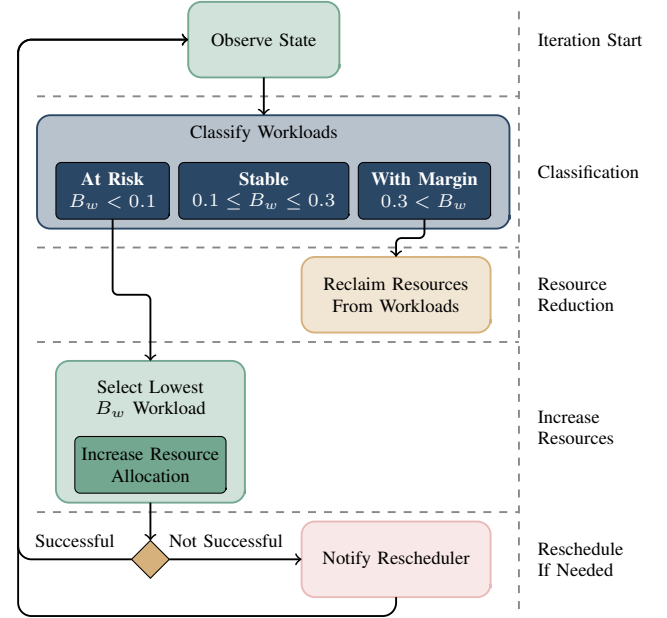

Our prototype controls two resources: CPU cores and last-level cache (LLC) capacity. 
The design is not inherently limited to these resources and can be extended to other 
resources for which suitable resource scores and allocation mechanisms are available.
The resource management controller periodically evaluates all workloads executing on 
its node. At each iteration, it obtains their current buoyancy and resource scores, 
classifies the workloads according to their performance headroom, and performs resource 
adjustments based on the resulting classification. Resource reductions may be performed 
for workloads with substantial headroom, while at most one workload experiencing 
resource pressure is given additional resources during each iteration. An overview
is shown in \figurename~\ref{fig:node-level}.

\subsubsection{Workload Classification}

As a first step, workloads are classified according to their buoyancy scores. 
We use three categories whose thresholds were experimentally selected to reflect 
the performance headroom of workloads:

\begin{itemize}
    \item \textbf{At risk:} Workloads with $B_w < 0.1$ are considered to be at risk of violating, or are already violating, their performance targets.
    \item \textbf{Stable:} Workloads with $0.1 \leq B_w \leq 0.3$ are considered to have limited performance headroom.
    \item \textbf{With margin:} Workloads with $B_w > 0.3$ are considered to have substantial performance headroom and are candidates for resource reduction.
\end{itemize}

The controller first attempts to reclaim resources from workloads classified
as \emph{with margin}. It then attempts to increase the resource allocation of 
the workload with the lowest buoyancy score among those classified as \emph{at risk}. 
Only one workload is given additional resources during each controller iteration. 
This gives the affected workload and the other workloads on the node time to 
respond to the allocation change before another increase is performed.
A parallel may be drawn to the additive increase and multiplicative 
decrease (AIMD) algorithm used in TCP congestion control~\cite{chiuAnalysisIncreaseDecrease1989}.

Limiting resource increases to one workload per iteration also reduces the 
likelihood of unnecessary allocation changes. Increasing the resources of one workload 
may change the resource pressure experienced by other workloads because they share 
the same underlying resources. Consequently, an increase that resolves the 
pressure of one workload may indirectly improve the buoyancy of other workloads.
If the controller cannot increase the resources available to the workload with 
the lowest buoyancy, local resource management is considered insufficient to 
resolve the observed resource pressure. The controller then signals the rescheduling component.
The resulting control loop is shown in Algorithm~\ref{alg:resource-management}.

\begin{algorithm}
\caption{Resource Management Controller}
\label{alg:resource-management}
\begin{algorithmic}[1]
\STATE $n \gets$ node hosting the resource management controller
\STATE $W \gets$ workloads on node $n$
\FOR{each iteration}
\STATE atRisk, stable, withMargin $\gets$ classifyWorkloads($W$)
\FOR{each workload $w$ in withMargin}
\STATE tryDecreaseResources($w$)
\ENDFOR
\STATE $w_{low} \gets$ lowestBuoyancyWorkload(atRisk)
\STATE increased $\gets$ \texttt{false}
\IF {$w_{low} \neq \texttt{null}$}
\STATE increased $\gets$ tryIncreaseResources($w_{low}$)
\ENDIF
\IF {not increased}
\STATE signalReschedulingComponent($n$)
\ENDIF
\ENDFOR
\end{algorithmic}
\end{algorithm}

\subsubsection{Resource Reduction}

Workloads classified as \emph{with margin} are candidates for resource reduction. 
The controller reduces at most one resource allocation for each workload during an 
iteration. The resource selected for reduction is the resource for which the 
workload's resource score has the smallest relative difference from 
the corresponding node-level resource score.
For each resource $r$, we define
\begin{equation}
\Delta_r =
\frac{R_{r,w} - R_{r,n}}{R_{r,n}}.
\end{equation}
The resource with the smallest $\Delta_r$ is selected for reduction, provided that 
the workload is not already at the minimum allocation for that resource. 
This comparison is intended to identify resources for which the workload is 
relatively less sensitive than the resource characteristics represented at the node level.
Importantly, the resource score represents workload sensitivity to a resource rather 
than its resource allocation or consumption directly. Consequently, the resource 
management policy uses resource scores to determine which allocation changes are likely 
to be least harmful rather than interpreting them as direct measures of resource demand.
The resource reduction procedure is shown in Algorithm~\ref{alg:try_decrease_resources}.

\begin{algorithm}
\caption{Subroutine: tryDecreaseResources}
\label{alg:try_decrease_resources}
\begin{algorithmic}[1]
\STATE $w \gets$ workload to decrease resources for
\STATE $n \gets$ node hosting workload $w$
\FOR {each resource $r$ in \{CPU, LLC\}}
\STATE $\Delta_{r} \gets
    \frac{R_{r,w} - R_{r,n}}{R_{r,n}}$
\ENDFOR
\STATE $r_{decrease} \gets$ resource with the lowest $\Delta_{r}$
\IF {$r_{decrease}$ is not at its minimum allocation}
    \STATE decreaseResourceAllocation($w$, $r_{decrease}$)
\ENDIF
\end{algorithmic}
\end{algorithm}

\subsubsection{Resource Increase}

When a workload is classified as \emph{at risk}, the workload with the lowest 
buoyancy score is selected for resource increase. The controller prioritizes 
resources according to the relative difference between the workload's 
resource score and the corresponding node-level resource score.
For each resource $r$, we compute
\begin{equation}
\Delta_r =
\frac{R_{r,w} - R_{r,n}}{R_{r,n}}.
\end{equation}
Resources are considered in descending order of $\Delta_r$. This prioritizes 
resources for which the workload exhibits greater sensitivity relative to the corresponding node-level resource score.

For each resource, the controller first checks whether unallocated capacity is 
available on the node. If sufficient capacity is available, the workload's 
allocation is increased directly. If no unallocated capacity is available, 
the controller attempts to reclaim part of the resource allocation from another workload on the node.
Candidate workloads are considered in ascending order of their resource score 
for the resource being allocated. A workload is eligible for resource reduction 
if it has a lower resource score than the workload requiring additional 
resources and is not already at its minimum possible allocation.\footnote{
    E.g., for LLC, the minimum possible allocation is a single cache way.
} 
This prioritizes transferring resources from workloads that are less sensitive to a resource to workloads more sensitive to it.

Because resource scores for a given resource are computed the same way on a given node, 
they can be directly compared when selecting candidate workloads. 
The complete resource-increase procedure is shown in Algorithm~\ref{alg:try_increase_resources}.

\begin{algorithm}
\caption{Subroutine: tryIncreaseResources}
\label{alg:try_increase_resources}
\begin{algorithmic}[1]
\STATE $w \gets$ workload to increase resources for
\STATE $n \gets$ node hosting workload $w$
\STATE $W$ $\gets$ workloads on node $n$
\FOR {each resource $r$ in \{CPU, LLC\}}
\STATE $\Delta_{r} \gets
    \frac{R_{r,w} - R_{r,n}}{R_{r,n}}$
\ENDFOR
\STATE tryIncreaseOrder $\gets$ resources sorted by $\Delta_{r}$ descending
\FOR {each resource $r$ in tryIncreaseOrder}
\IF {unallocated resources of $r$ are available}
    \STATE increaseResourceAllocation($w$, $r$)
    \RETURN \texttt{true}
\ENDIF 
\STATE $W_{asc} \gets$ workloads in $W$ sorted by $R_{r,w}$ ascending
\FOR {each workload $w_{other}$ in $W_{asc}$}
    \IF {$w_{other} \neq w$ and $R_{r,w_{other}} < R_{r,w}$}
    \IF {$w_{other}$ is not at minimum allocation of $r$} 
        \STATE decreaseResourceAllocation($w_{other}$, $r$)
        \STATE increaseResourceAllocation($w$, $r$)
        \RETURN \texttt{true}
    \ENDIF
    \ENDIF
\ENDFOR
\ENDFOR
\RETURN \texttt{false}
\end{algorithmic}
\end{algorithm}

\subsection{Rescheduling Component} \label{sec:rescheduling}

The rescheduling component provides a mechanism for resolving resource pressure 
that cannot be addressed through local resource allocation. It is triggered 
when the resource management component is unable to provide additional resources to a workload experiencing low buoyancy.

When triggered, the rescheduling component evaluates the workloads executing 
on the affected node and selects one workload to move away from the node. 
The selected workload is removed from the node and returned to the orchestrator's 
scheduling process. The workload can then be placed on another node using the current buoyancy characteristics of the cluster.

The workload selected for rescheduling is the workload with the lowest buoyancy 
score on the affected node. This prioritizes workloads with the least performance 
headroom and provides an opportunity to relieve the pressure experienced by 
the workload most affected by the current placement.

Only one workload is rescheduled in response to a resource management event. 
Following the migration, the resource management component is given time 
to observe the resulting workload composition and resource characteristics 
before another rescheduling action is initiated. This prevents repeated rescheduling 
actions in response to transient changes in resource pressure and allows the 
effects of a placement change to propagate through the node.

The rescheduling component is logically distinct from the resource management 
component, even though the two operate closely together. The resource management 
component is responsible for deciding whether local resource allocation is sufficient, 
while the rescheduling component is responsible for resolving situations where 
local allocation is insufficient.

%% file: resources/tikz/approach.tikz
\begin{tikzpicture}
	\begin{pgfonlayer}{nodelayer}
		\node [style=Small Circle] (0) at (-3, 0) {};
		\node [style=none] (2) at (-0.75, 0.75) {};
		\node [style=none] (3) at (-0.75, -0.75) {};
		\node [style=none] (4) at (2.25, -0.75) {};
		\node [style=none] (5) at (2.25, 0.75) {};
		\node [style=none] (7) at (4.25, 0.75) {};
		\node [style=none] (8) at (4.25, -0.75) {};
		\node [style=none] (9) at (7.25, -0.75) {};
		\node [style=none] (10) at (7.25, 0.75) {};
		\node [style=none] (13) at (-1.25, 1.75) {};
		\node [style=none] (14) at (-1.25, -1) {};
		\node [style=none] (15) at (7.75, -1) {};
		\node [style=none] (16) at (7.75, 1.75) {};
		\node [style=none] (17) at (-0.75, 0) {};
		\node [style=none] (18) at (2.25, 0) {};
		\node [style=none] (19) at (4.25, 0) {};
		\node [style=none] (20) at (7.25, 0) {};
		\node [style=none] (21) at (3.25, 0.5) {};
		\node [style=none] (22) at (2.75, 0) {};
		\node [style=none] (23) at (3.75, 0) {};
		\node [style=none] (24) at (3.25, -0.5) {};
		\node [style=none] (25) at (3.25, -1.5) {};
		\node [style=none] (26) at (-2, -1.5) {};
		\node [style=none] (27) at (-2, 0) {};
		\node [style=Small Circle] (28) at (9.5, 0) {};
		\node [style=Small Circle, fill={rgb,255: red,115; green,167; blue,144}] (30) at (3.25, -3) {};
		\node [style=none] (31) at (9.5, -1.5) {};
		\node [style=none] (32) at (3.25, -1.5) {};
		\node [style=none] (33) at (-1.25, -2) {};
		\node [style=none] (34) at (-1.25, -5.25) {};
		\node [style=none] (35) at (7.75, -5.25) {};
		\node [style=none] (36) at (7.75, -2) {};
		\node [style=none] (38) at (-0.75, -3.5) {};
		\node [style=none] (39) at (-0.75, -5) {};
		\node [style=none] (40) at (2.75, -5) {};
		\node [style=none] (41) at (2.75, -3.5) {};
		\node [style=none] (42) at (3.75, -3.5) {};
		\node [style=none] (43) at (3.75, -5) {};
		\node [style=none] (44) at (7.25, -5) {};
		\node [style=none] (45) at (7.25, -3.5) {};
		\node [style=none] (46) at (1, -3.5) {};
		\node [style=none] (47) at (1, -3) {};
		\node [style=none] (48) at (2, -3) {};
		\node [style=none] (49) at (5.5, -3.5) {};
		\node [style=none] (51) at (4.5, -3) {};
		\node [style=none] (52) at (5.5, -3) {};
		\node [style=none] (53) at (2.75, -4.25) {};
		\node [style=none] (54) at (3.75, -4.25) {};
		\node [style=none] (57) at (-1.25, -6.25) {};
		\node [style=none] (58) at (-1.25, -9) {};
		\node [style=none] (59) at (7.75, -9) {};
		\node [style=none] (60) at (7.75, -6.25) {};
		\node [style=none] (62) at (1.5, -7.25) {};
		\node [style=none] (63) at (1.5, -8.75) {};
		\node [style=none] (64) at (5, -8.75) {};
		\node [style=none] (65) at (5, -7.25) {};
		\node [style=none] (67) at (5.5, -5) {};
		\node [style=none] (68) at (5.5, -5.75) {};
		\node [style=none] (69) at (3.25, -5.75) {};
		\node [style=none] (70) at (3.25, -7.25) {};
		\node [style=none] (71) at (3.25, -8.75) {};
		\node [style=none] (72) at (3.25, -9.25) {};
		\node [style=none] (73) at (-2, -9.25) {};
		\node [style=none] (74) at (-0.75, 0) {};
		\node [style=none] (75) at (-2, 0) {};
		\node [style=Text, font={\scriptsize}] (1) at (-3, 1) {New Arrival};
		\node [style=Text, font={\scriptsize}, anchor=west] (12) at (-1, 1.25) {Scheduling};
		\node [style=Text, font={\scriptsize}] (29) at (9.5, 1.25) {Node\\Assignment};
		\node [style=Text, font={\scriptsize}] (6) at (0.75, 0) {Filtering};
		\node [style=Text, font={\scriptsize}] (11) at (5.75, 0) {Scoring};
		\node [style=Text, font={\scriptsize}, anchor=west] (37) at (-1, -2.5) {Resource Mgmt};
		\node [style=White Text, font={\scriptsize}] (55) at (1, -4.25) {Monitor};
		\node [style=White Text, font={\scriptsize}] (56) at (5.5, -4.25) {Allocation\\Adjustment};
		\node [style=Text, font={\scriptsize}, anchor=west] (61) at (-1, -6.75) {Rescheduling};
		\node [style=Text, font={\scriptsize}] (66) at (3.25, -8) {Eviction};
	\end{pgfonlayer}
	\begin{pgfonlayer}{edgelayer}
		\draw [style=Blue EC F, rounded corners=2pt] (14.center)
			 to (15.center)
			 to (16.center)
			 to (13.center)
			 to cycle;
		\draw [style=Freebox Green, rounded corners=5pt] (3.center)
			 to (4.center)
			 to (5.center)
			 to (2.center)
			 to cycle;
		\draw [style=Freebox Green, rounded corners=5pt] (8.center)
			 to (9.center)
			 to (10.center)
			 to (7.center)
			 to cycle;
		\draw [style=Freebox Yellow] (24.center)
			 to (23.center)
			 to (21.center)
			 to (22.center)
			 to cycle;
		\draw [style=Green EC F, rounded corners=2pt] (34.center)
			 to (35.center)
			 to (36.center)
			 to (33.center)
			 to cycle;
		\draw [style=Freebox, rounded corners=5pt] (39.center)
			 to (40.center)
			 to (41.center)
			 to (38.center)
			 to cycle;
		\draw [style=Freebox, rounded corners=5pt] (43.center)
			 to (44.center)
			 to (45.center)
			 to (42.center)
			 to cycle;
		\draw [style=Yellow EC F, rounded corners=2pt] (58.center)
			 to (59.center)
			 to (60.center)
			 to (57.center)
			 to cycle;
		\draw [style=Freebox Red, rounded corners=5pt] (63.center)
			 to (64.center)
			 to (65.center)
			 to (62.center)
			 to cycle;
		\draw [style=Arrow] (0) to (17.center);
		\draw [style=Arrow] (18.center) to (22.center);
		\draw [style=Arrow] (23.center) to (19.center);
		\draw [style=Dashed Arrow, rounded corners=5pt] (24.center)
			 to (25.center)
			 to (26.center)
			 to (27.center)
			 to (17.center);
		\draw [style=Arrow] (20.center) to (28);
		\draw [style=Arrow, rounded corners=5pt] (28)
			 to (31.center)
			 to (32.center)
			 to (30);
		\draw [style=Arrow, rounded corners=2pt] (30)
			 to (48.center)
			 to (47.center)
			 to (46.center);
		\draw [style=Arrow, rounded corners=2pt] (49.center)
			 to (52.center)
			 to (51.center)
			 to (30);
		\draw [style=Arrow] (53.center) to (54.center);
		\draw [style=Dashed Arrow, rounded corners=5pt] (67.center)
			 to (68.center)
			 to (69.center)
			 to (70.center);
		\draw [style=Dashed Arrow, rounded corners=5pt] (71.center)
			 to (72.center)
			 to (73.center)
			 to (75.center)
			 to (74.center);
	\end{pgfonlayer}
\end{tikzpicture}

%% file: resources/tikz/resource-manager.tikz
\begin{tikzpicture}
	\begin{pgfonlayer}{nodelayer}
		\node [style=none] (2) at (-3, 1.25) {};
		\node [style=none] (3) at (-3, -0.75) {};
		\node [style=none] (4) at (1, -0.75) {};
		\node [style=none] (5) at (1, 1.25) {};
		\node [style=none] (7) at (-7, -1.75) {};
		\node [style=none] (8) at (-7, -4.75) {};
		\node [style=none] (9) at (5.5, -4.75) {};
		\node [style=none] (10) at (5.5, -1.75) {};
		\node [style=none] (12) at (-6.5, -3) {};
		\node [style=none] (13) at (-6.5, -4.5) {};
		\node [style=none] (14) at (-3.5, -4.5) {};
		\node [style=none] (15) at (-3.5, -3) {};
		\node [style=none] (17) at (-3.25, -3) {};
		\node [style=none] (18) at (-3.25, -4.5) {};
		\node [style=none] (19) at (1.25, -4.5) {};
		\node [style=none] (20) at (1.25, -3) {};
		\node [style=none] (22) at (1.5, -3) {};
		\node [style=none] (23) at (1.5, -4.5) {};
		\node [style=none] (24) at (5, -4.5) {};
		\node [style=none] (25) at (5, -3) {};
		\node [style=none] (27) at (0, -5.5) {};
		\node [style=none] (28) at (0, -7.5) {};
		\node [style=none] (29) at (5, -7.5) {};
		\node [style=none] (30) at (5, -5.5) {};
		\node [style=none] (32) at (-6.5, -8.25) {};
		\node [style=none] (33) at (-6.5, -12) {};
		\node [style=none] (34) at (-1.5, -12) {};
		\node [style=none] (35) at (-1.5, -8.25) {};
		\node [style=none] (37) at (-6, -10.25) {};
		\node [style=none] (38) at (-6, -11.75) {};
		\node [style=none] (39) at (-2, -11.75) {};
		\node [style=none] (40) at (-2, -10.25) {};
		\node [style=none] (42) at (-4, -13) {};
		\node [style=none] (43) at (-4.5, -13.5) {};
		\node [style=none] (44) at (-3.5, -13.5) {};
		\node [style=none] (45) at (-4, -14) {};
		\node [style=none] (46) at (0, -12.5) {};
		\node [style=none] (47) at (0, -14.5) {};
		\node [style=none] (48) at (5, -14.5) {};
		\node [style=none] (49) at (5, -12.5) {};
		\node [style=none] (51) at (-1, -0.75) {};
		\node [style=none] (52) at (-1, -1.75) {};
		\node [style=none] (53) at (3.25, -4.5) {};
		\node [style=none] (54) at (3.25, -5) {};
		\node [style=none] (55) at (2.5, -5) {};
		\node [style=none] (56) at (2.5, -5.5) {};
		\node [style=none] (57) at (-5, -4.5) {};
		\node [style=none] (58) at (-5, -6.5) {};
		\node [style=none] (59) at (-4, -6.5) {};
		\node [style=none] (60) at (-4, -8.25) {};
		\node [style=none] (61) at (-4, -12) {};
		\node [style=none] (62) at (2.5, -14.5) {};
		\node [style=none] (63) at (2.5, -15) {};
		\node [style=none] (64) at (-4, -15) {};
		\node [style=none] (65) at (-7.5, -15) {};
		\node [style=none] (66) at (-7.5, 0.25) {};
		\node [style=none] (67) at (-3, 0.25) {};
		\node [style=none] (68) at (0, -13.5) {};
		\node [style=none] (69) at (-7.5, -13.5) {};
		\node [style=none] (72) at (-7, -1.25) {};
		\node [style=none] (73) at (5.75, -1.25) {};
		\node [style=none] (74) at (-7, -5.25) {};
		\node [style=none] (75) at (5.75, -5.25) {};
		\node [style=none] (76) at (-7, -7.75) {};
		\node [style=none] (77) at (5.75, -7.75) {};
		\node [style=none] (78) at (5.75, 1) {};
		\node [style=none] (79) at (5.75, -15) {};
		\node [style=none] (83) at (-7, -12.25) {};
		\node [style=none] (84) at (5.75, -12.25) {};
		\node [style=Text, font={\scriptsize}] (6) at (-1, 0.25) {Observe State};
		\node [style=Text, font={\scriptsize}, anchor=west] (1) at (6, 0.25) {Iteration Start};
		\node [style=Text, font={\scriptsize}] (11) at (-1, -2.25) {Classify Workloads};
		\node [style=White Text, font={\scriptsize}] (16) at (-5, -3.75) {\textbf{At Risk}\\$B_w < 0.1$};
		\node [style=White Text, font={\scriptsize}] (21) at (-1, -3.75) {\textbf{Stable}\\$0.1 \leq B_w \leq 0.3$};
		\node [style=White Text, font={\scriptsize}] (26) at (3.25, -3.75) {\textbf{With Margin}\\$0.3 < B_w$};
		\node [style=Text, font={\scriptsize}, anchor=west] (80) at (6, -3.25) {Classification};
		\node [style=Text, font={\scriptsize}] (31) at (2.5, -6.5) {Reclaim Resources\\From Workloads};
		\node [style=Text, font={\scriptsize}, anchor=west, align=left] (81) at (6, -6.5) {Resource\\Reduction};
		\node [style=Text, font={\scriptsize}] (36) at (-4, -9.25) {Select Lowest\\$B_w$ Workload};
		\node [style=Text, font={\scriptsize}] (41) at (-4, -11) {Increase Resource\\Allocation};
		\node [style=Text, font={\scriptsize}, anchor=west, align=left] (82) at (6, -10) {Increase\\Resources};
		\node [style=Text, font={\scriptsize}] (70) at (-6, -13) {Successful};
		\node [style=Text, font={\scriptsize}] (71) at (-2, -13) {Not Successful};
		\node [style=Text, font={\scriptsize}] (50) at (2.5, -13.5) {Notify Rescheduler};
		\node [style=Text, font={\scriptsize}, anchor=west, align=left] (85) at (6, -13.5) {Reschedule\\If Needed};
	\end{pgfonlayer}
	\begin{pgfonlayer}{edgelayer}
		\draw [style=Freebox Yellow] (44.center)
			 to (45.center)
			 to (43.center)
			 to (42.center)
			 to cycle;
		\draw [style=Green EC F, rounded corners=5pt] (4.center)
			 to (5.center)
			 to (2.center)
			 to (3.center)
			 to cycle;
		\draw [style=Blue EC F, rounded corners=5pt] (9.center)
			 to (10.center)
			 to (7.center)
			 to (8.center)
			 to cycle;
		\draw [style=Freebox, rounded corners=2pt] (14.center)
			 to (15.center)
			 to (12.center)
			 to (13.center)
			 to cycle;
		\draw [style=Freebox, rounded corners=2pt] (19.center)
			 to (20.center)
			 to (17.center)
			 to (18.center)
			 to cycle;
		\draw [style=Freebox, rounded corners=2pt] (24.center)
			 to (25.center)
			 to (22.center)
			 to (23.center)
			 to cycle;
		\draw [style=Yellow EC F, rounded corners=5pt] (29.center)
			 to (30.center)
			 to (27.center)
			 to (28.center)
			 to cycle;
		\draw [style=Green EC F, rounded corners=5pt] (34.center)
			 to (35.center)
			 to (32.center)
			 to (33.center)
			 to cycle;
		\draw [style=Freebox Green, rounded corners=2pt] (39.center)
			 to (40.center)
			 to (37.center)
			 to (38.center)
			 to cycle;
		\draw [style=Pink EC F, rounded corners=5pt] (48.center)
			 to (49.center)
			 to (46.center)
			 to (47.center)
			 to cycle;
		\draw [style=Dashed] (72.center) to (73.center);
		\draw [style=Dashed] (74.center) to (75.center);
		\draw [style=Dashed] (76.center) to (77.center);
		\draw [style=Dashed] (78.center) to (79.center);
		\draw [style=Dashed] (83.center) to (84.center);
		\draw [style=Arrow] (51.center) to (52.center);
		\draw [style=Arrow, rounded corners=2pt] (57.center)
			 to (58.center)
			 to (59.center)
			 to (60.center);
		\draw [style=Arrow, rounded corners=2pt] (53.center)
			 to (54.center)
			 to (55.center)
			 to (56.center);
		\draw [style=Arrow] (61.center) to (42.center);
		\draw [style=Arrow] (44.center) to (68.center);
		\draw [style=Arrow, rounded corners=5pt] (62.center)
			 to (63.center)
			 to (64.center)
			 to (65.center)
			 to (66.center)
			 to (67.center);
		\draw [style=Arrow, rounded corners=5pt] (43.center)
			 to (69.center)
			 to (66.center)
			 to (67.center);
	\end{pgfonlayer}
\end{tikzpicture}

%% file: content/50.implementation.tex
\section{Implementation Details}\label{sec:implementation}

The cloud native ecosystem, centered around 
Kubernetes,\footnote{\url{https://kubernetes.io/}} has become a 
widely adopted platform for deploying and managing applications in cloud 
environments. While the solution design is independent of any particular 
platform or orchestrator, we implement our approach as a 
Kubernetes-native system to leverage Kubernetes' extensibility and 
existing workload-management mechanisms.

The three logical components are implemented using different Kubernetes 
mechanisms according to their respective responsibilities. 
The scheduling component is implemented as a Kubernetes scheduler 
plugin, the resource management component as a node-local Kubernetes 
controller deployed as a DaemonSet, and the rescheduling logic as 
part of the resource management controller.

\subsection{Kubernetes Integration}

\subsubsection{Scheduling}

The scheduling component is implemented as a Kubernetes scheduler plugin. 
The plugin implements the filtering and scoring policies described in 
Section~\ref{sec:scheduling-design} and integrates them directly into the Kubernetes scheduling process.

During scheduling, the plugin is invoked for candidate nodes and 
determines whether each node satisfies the buoyancy-based filtering criterion. 
For feasible nodes, the plugin calculates a score based on the node's buoyancy. 
These scores are then provided to the Kubernetes scheduler, which combines them 
with the scores of other configured scheduling plugins according to the 
scheduler configuration~\cite{larsson_impact_2023}.

Kubernetes expects scheduler plugin scores to be represented as positive integers. 
We therefore scale the buoyancy-based score 
as described in Section~\ref{sec:scoring} to the range compatible with 
the Kubernetes scheduler. The scaling does not change the ordering of candidate nodes and 
consequently does not alter the underlying scheduling policy.

\subsubsection{Resource Management}

The resource management component is implemented as a Kubernetes controller. 
A separate controller instance runs on each worker node and manages 
the resource allocations of the Pods executing on that node.
The controller periodically queries the current state of the Pods and their 
corresponding buoyancy and resource scores. It then applies the control policy 
described in Section~\ref{sec:resource-management}. Because the controller operates 
independently on each node, resource-management decisions are based on 
the local workload composition and resource state rather than requiring 
a centralized resource-management controller.

\subsubsection{Rescheduling}

The rescheduling component is logically separate from resource management 
but is integrated into the node-local resource management controller in 
our implementation. This integration is motivated by the fact that 
rescheduling decisions are triggered by node-local resource pressure and 
use the same workload and resource state already collected by the resource management controller.

When the resource management controller determines that local resource allocation 
cannot resolve the resource pressure, it invokes the rescheduling 
logic. The controller selects a workload according to the 
policy described in Section~\ref{sec:rescheduling} and deletes the corresponding Pod through the Kubernetes API.

For workloads managed by a higher-level Kubernetes controller, such as 
a Deployment or StatefulSet, deletion of the Pod causes the higher-level 
controller to create a replacement Pod. The replacement Pod is subsequently placed 
through the normal Kubernetes scheduling process and is therefore subject to the buoyancy-aware scheduling plugin.

This implementation effectively closes the feedback loop between rescheduling and 
scheduling: resource pressure can cause a workload to leave a node, 
after which the replacement workload is placed using the current state of the cluster.

\subsection{Dynamic CPU Allocation}

We use CPU core pinning together with dynamic CPU allocation to provide predictable 
CPU resource isolation while allowing allocations to change at runtime. 
CPU allocation is implemented using the \emph{static} policy of the 
Kubernetes CPU Manager together with the \emph{InPlacePodVerticalScaling} feature.

At the time of writing, the static CPU Manager policy 
and \emph{InPlacePodVerticalScaling} are incompatible in the upstream Kubernetes 
release. We therefore use a patched version of Kubernetes based on an upstream 
pull request that enables the two features to operate together~\cite{inplacepodverticalscaling}.

The resource management controller modifies Pod CPU allocations through 
the Kubernetes Pod \emph{resize} sub-resource. The CPU Manager then handles 
the corresponding allocation of CPU cores to the Pod. Because CPU allocations 
are expressed in integer numbers of cores when using the static CPU Manager 
policy, each allocation change corresponds to an increase or decrease of at least one full CPU core.
CPU pinning ensures that allocated cores remain dedicated to the corresponding Pod. 
This provides stronger CPU isolation than relying solely on proportional CPU scheduling 
and is important for reducing interference and improving the predictability of the resource-management actions.

\subsection{Dynamic LLC Allocation}

We manage LLC allocation using Intel Resource Director 
Technology (RDT) and the \emph{resctrl} file system~\cite{yuUserInterfaceResource}. 
Intel RDT provides cache allocation through \emph{class of service} (COS) 
configurations. A COS specifies an LLC allocation policy, and processes 
are assigned to a COS to enforce the corresponding allocation.
Because Kubernetes does not natively expose the required 
\emph{resctrl} functionality, we implement a 
node-local \emph{resctrl agent} as a DaemonSet. The agent monitors 
Kubernetes annotations and translates them into \emph{resctrl} configurations. 
A node annotation defines the LLC allocation associated with each COS, 
while a Pod annotation determines the COS to which the Pod's container processes are assigned.

The number of available COSs and LLC cache ways is limited by the underlying 
hardware and varies between processor generations and 
models~\cite{larssonHardwareLevelQoSEnforcement2025}. Consequently, assigning a 
dedicated COS to every workload is not generally possible. 
Our implementation therefore assigns a dedicated COS only to workloads classified 
as \emph{at risk}. Workloads in the other states remain in the default 
COS and share cache ways that have not been exclusively allocated.
Once a Pod is assigned to a dedicated COS, it remains associated with that 
COS until it is evicted from the node. This ensures that the cache 
allocation remains stable between resource-management iterations and avoids repeatedly changing 
the underlying \emph{resctrl} configuration.

%% file: content/55.experimental-setup.tex
\section{Experimental Setup}\label{sec:experimental-platform}

We evaluate our approach using a Kubernetes cluster deployed on the Grid'5000 testbed 
infrastructure.\footnote{\url{https://www.grid5000.fr/}} 
On this hardware, we deploy a cluster configured 
to emulate a cloud environment representative of a 
real-world deployment. The cluster comprises 11 nodes 
with identical hardware configurations, as summarized 
in \tablename~\ref{tab:platform}.

\begin{table}[tbp]
    \centering
    \caption{Experimental platform node hardware configuration.}
    \footnotesize
    \begin{tabular}{ll}
        \toprule
        \textbf{Component} & \textbf{Specification} \\ \midrule
        
        \textbf{CPU Model} & Intel Xeon Gold 5320 \\
        \textbf{Operating System} & Debian 12.2 (kern. 6.1) \\ 
        \textbf{Sockets} & 2 \\
        \textbf{CPU Cores (Physical / Logical)} & 52 / 104 \\
        \textbf{Frequency (Base / Max Turbo)} & 2.2 GHz / 3.4 GHz \\
        \textbf{L1 Cache per Core (Inst / Data)} & 48 KiB / 32 KiB \\
        \textbf{L2 Cache per Core} & 1280 KiB \\
        \textbf{LLC per Socket} & 39 MiB, 12 ways \\
        \textbf{Memory (Capacity / Speed)} & 384 GiB / 2933 MT/s \\
        \textbf{Memory (Channels / Bus Width)} & 6 / 8 Bytes \\
        \textbf{Storage} & NVME SSD \\
        \textbf{Network} & 25 Gbps \\
        \bottomrule
    \end{tabular}
    \label{tab:platform}
\end{table}

The nodes are assigned to three roles: one node 
serves as the Kubernetes control plane, five worker nodes 
host the application workloads, and the remaining five nodes 
generate application load. This separation of concerns
ensure there is no additional resource contention created locally on
the nodes under test from load generation or control-plane operations.
Each worker node runs a local 
buoyancy agent that monitors resource usage and computes resource 
and buoyancy scores for both the node and the workloads running on it. 
A central Prometheus\footnote{\url{https://prometheus.io/}} 
time-series database runs on the control-plane node and collects 
and aggregates the metrics required by the scheduling and rescheduling 
components.

\subsection{Workload Selection}\label{sec:workload-selection}

To understand how our solution performs when controlling a cluster hosting
workloads of different characteristics, we select a set of workloads 
that represent a diverse range of resource usage patterns. 
\tablename~\ref{tab:workloads} lists the workloads used in our evaluation.
Nginx is a web server that is widely used in cloud environments and represent
a common lightweight cloud workload. Moses, img-dnn and xapian are 
part of the TailBench benchmark 
suite~\cite{kastureTailbenchBenchmarkSuite2016} and represent more 
resource-intensive workloads.

\begin{table}[tbp]
    \centering
    \caption{Workloads used in the evaluation.}
    \footnotesize
    \begin{tabular}{ll}
        \toprule
        \textbf{Workload} & \textbf{Description} \\ \midrule
        \texttt{nginx} & Web server \\
        \texttt{moses} & Statistical machine translation engine \\
        \texttt{img-dnn} & Deep neural network image classification \\
        \texttt{xapian} & Search engine \\
        \bottomrule
    \end{tabular}
    \label{tab:workloads}
\end{table}
 
All load generation is performed from the load generator nodes using a custom
open-loop~\cite{schroederOpenClosedCautionary2006} load generator that collects and exposes metrics which are 
then aggregated by Prometheus. Each workload Pod is deployed 
alongside one or more load generator instances whose load and reported 
metrics only correspond to that that specific workload instance.

Nginx serves $10^6$ static HTML pages, each 1KiB in size. These pages
are queried using a Zipf distribution. Moses, img-dnn, and xapian
from the open-source Tailbench suite were converted into long-running services and containerized.
Load for these workloads was generated using the same configuration as
in the original benchmark with respect to input data, request distribution and problem space sizing.

Sizing experiments were performed to determine appropriate performance targets
for each workload. These were carried out by deploying a workload in isolation,
allocating it 6 cores, and varying the load over time to
identify a ``knee'' in the performance curve similar to previous 
studies~\cite{chen_parties_2019, patel_clite_2020, larsson_beyond_2026}.
The load was set to $80\%$ of the knee load and the SLO was 
set to $150\%$ of the $99^\text{th}$ percentile tail latency measured
at the knee load. The results of the sizing experiments are summarized in \tablename~\ref{tab:loads}.
It is worth noting that this information is strictly used for experiment
configuration and sizing and no profiling or prior knowledge of the workloads 
is used by the proposed resource management mechanisms.

\begin{table}[tbp]
    \centering
    \caption{Results of the workload sizing experiments.}
    \footnotesize
    \begin{tabular}{lll}
        \toprule
        \textbf{Workload} & \textbf{Base Load (RPS)} & \textbf{P99 SLO (ms)} \\ \midrule
        \texttt{nginx} & $40000$ & $41.87$ \\
        \texttt{moses} & $1360$ & $14.31$ \\
        \texttt{img-dnn} & $1440$ & $14.72$ \\
        \texttt{xapian} & $1200$ & $27.62$ \\
        \bottomrule
    \end{tabular}
    \label{tab:loads}
\end{table}

%% file: content/60.evaluation.tex
\section{Evaluation}

In the process of evaluating the effectiveness of our solution, we ask a series
of questions that we answer experimentally:

\begin{enumerate}
    \item How does the proposed solution affect the performance of workloads across 
        a dynamic Kubernetes environment?
    \item Is the resource utilization of a cluster improved by our proposed solution?
    \item Is our solution able to reduce resource fragmentation across a cluster? \item How large are the rescheduling overheads introduced by our solution?
\end{enumerate}

To answer these questions, we start by designing experiments as described in Section~\ref{sec:experiment-design}. 
We then answer questions by examining the results of our experiments in Section~\ref{sec:experiment-results}.

\subsection{Experiment Design}\label{sec:experiment-design}

We design our experiments on top of the Kubernetes-based platform 
described in Section~\ref{sec:experimental-platform}. All experiments
are first run on a default Kubernetes configuration without our solution 
enabled to gather a baseline. They are later repeated with our solution enabled 
to evaluate its impact on the performance of workloads, resource utilization, 
resource fragmentation, and rescheduling overheads. All experiments are
controlled by an automated script that is executed on the cluster's control plane node.

Our main experiment consists of repeatedly deploying workloads at set time intervals.
Every 15 seconds, 5 workloads (matching the number of worker nodes) are randomly selected from the set of workloads
described in Section~\ref{sec:workload-selection} and deployed to the cluster alongside
their respective load generators. 
Once created, we let the Kubernetes attempt to place the workloads onto suitable worker nodes. 
Should no node be available for a workload and scheduling fails, 
the workload is simply discarded and deleted. This workload creation process continues for 10 minutes.
Once stopped, the cluster is given 3 minutes to stabilize before cluster state 
information is gathered and performance metrics are collected. By repeating
this experiment, we can ensure a varied and unpredictable workload composition on the cluster,
which allows us to evaluate the performance of our solution under different conditions and workload compositions.

\subsection{Experimental Results and Discussion}\label{sec:experiment-results}

This section presents the results of our experiments and discusses their implications.
All results are collected from 25 repetitions for each configuration.

\subsubsection{Workload Performance Impact}

\begin{figure}[tbp]
    \centering
    \scalebox{.5}{
        \fontsize{14}{14}\selectfont 
        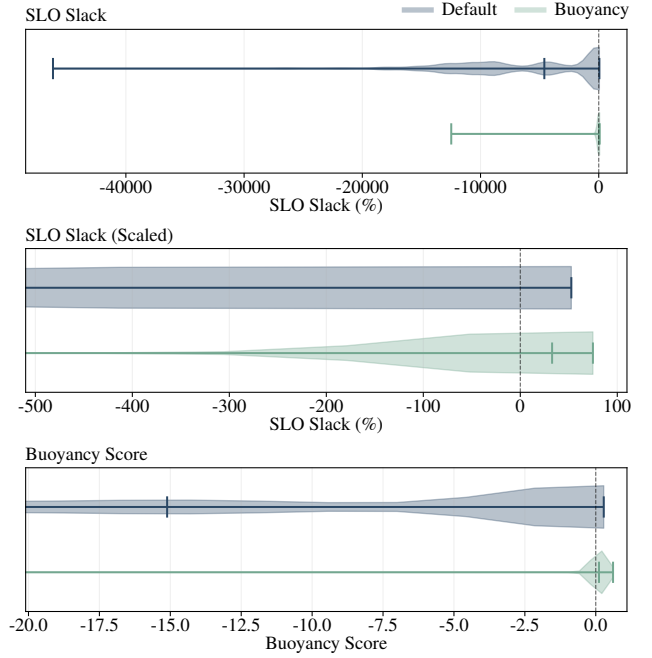
    }
    \caption{SLO slack percentages and buoyancy scores 
        for workload deployed in the cluster. It is clear that our buoyancy-based
        solution reduces the number of workloads that are in violation of their SLOs 
        while also reducing the size of any violations that do occur.
    }
    \label{fig:slo-buoyancy}
\end{figure}

We first examine the performance results of workloads deployed in the cluster, 
presented in \figurename~\ref{fig:slo-buoyancy}. 
The figure shows the SLO slack percentages and buoyancy scores for workloads deployed in the cluster.
The SLO slack percentage is a generalized metric that indicates how well a workload is performing relative to its SLO.
If $K_{actual}$ is the current value of a workload KPI metric and $K_{SLO}$ is the target value of that KPI,
then the SLO slack percentage is given by
\begin{equation}
    \frac{K_{SLO} - K_{actual}}{K_{SLO}},
\end{equation}
where a positive value indicates that the workload is performing better than its SLO and has slack to spare, 
while a negative value indicates that the workload is performing worse than its SLO and is in violation.

The results from \figurename~\ref{fig:slo-buoyancy} show that our buoyancy based solution keeps
most of the workloads within their performance targets as indicated by a positive SLO slack percentage.
Still, some workloads experience SLO violations, which is expected given the 
dynamic nature of the workload composition and their dynamic load patterns.
It is notable that the size of the SLO violations is much greater in the default 
configuration than in our buoyancy-based configuration. Thus, in answer to question 1,
we conclude that our solution is able to improve the performance of workloads in a dynamic Kubernetes environment.

\subsubsection{Resource Utilization Impact}

\begin{figure}[tbp]
    \centering
    \scalebox{.5}{
        \fontsize{14}{14}\selectfont 
        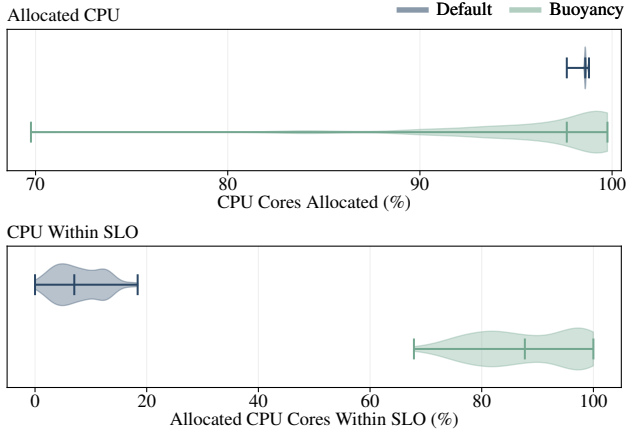
    }
    \caption{The percentage of available CPU cores on worker nodes allocated 
        to workloads and the percentage of currently allocated cores that are assigned to 
        workloads that are compliant with their SLOs at the time of measurement.
    }
    \label{fig:allocated-compliance}
\end{figure}

To understand the impact of our solution on resource utilization, 
we examine the allocation results presented in \figurename~\ref{fig:allocated-compliance}.
We see that the default configuration, which does not have any rescheduling mechanisms, is able to consistently
allocate close to $100\%$ of the available CPU cores on the worker nodes. This is to be expected as Kubernetes
will admit workloads as long as there are CPU and memory resources available at the time of scheduling.
Our solution is not as consistent, but it is able to allocate a similar amount of CPU cores on average.
It is notable that some data points show a lower allocation percentage, which is due to the fact that our 
solution may evict workloads from nodes to improve the performance of other workloads if it could not do so by 
locally adjusting resource allocations. This is a trade-off that our solution makes to ensure 
that workloads are able to meet their SLOs and these cores may be allocated to other workloads that are more compatible 
with the node's current resource congestion. Even so, the percentage of those cores that are assigned to workloads
running within their SLOs is much higher in our solution than in the default configuration, 
$87.8\%$ compared to $7.74\%$ in the default configuration on average. We also recall the fact
that the size of the SLO violations is much greater in the default configuration 
than in our buoyancy-based configuration. To answer question 2, the results indicate
a small decrease in resource utilization as measured by allocated CPU cores, 
but a significant increase in the percentage of those allocated cores 
that are assigned to workloads that are compliant with their SLOs.

\subsubsection{Resource Fragmentation Effects}

\begin{figure*}[tbp]
    \centering
    \scalebox{.5}{
        \fontsize{14}{14}\selectfont 
        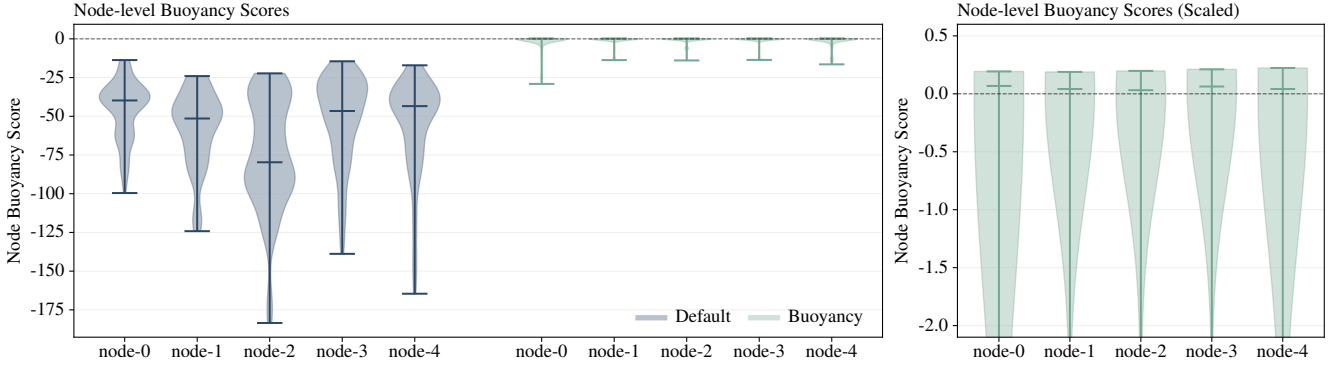
    }
    \caption{Node-level buoyancy scores for each worker node in the cluster. The node-level buoyancy score
    is a weighted average of the buoyancy scores of all workloads running on the node, 
    where the workload buoyancies are computed from the resource scores and application-level metrics of each workload.}
    \label{fig:node-buoyancy}
\end{figure*}

Previous studies have shown that fragmentation can happen in Kubernetes
clusters over time and that this can lead to stranded resources and poor resource utilization~\cite{larsson_impact_2023}.
Our solution is designed to counter the effect through resource insights and rescheduling actions.
By observing the node-level buoyancy scores presented in \figurename~\ref{fig:node-buoyancy},
we can gain a high-level understanding of the resource pressures that exist on each worker node in the cluster.

We see that the default configuration displays much larger variance in node-level 
buoyancy scores than our solution, which indicates that some nodes 
are under high resource pressure while others are under comparatively low pressure. 
This is a clear indication that workloads that compete over the same resources 
are being scheduled onto the same nodes, leading to severe resource congestion and cluster fragmentation.
The node-level buoyancy scores observed for our solution are much more consistent across nodes
and iterations, which indicates that workloads are better distributed across the cluster 
according to their resource requirements and the current resource pressures on each node.

\begin{figure*}[tbp]
    \centering
    \scalebox{.5}{
        \fontsize{14}{14}\selectfont 
        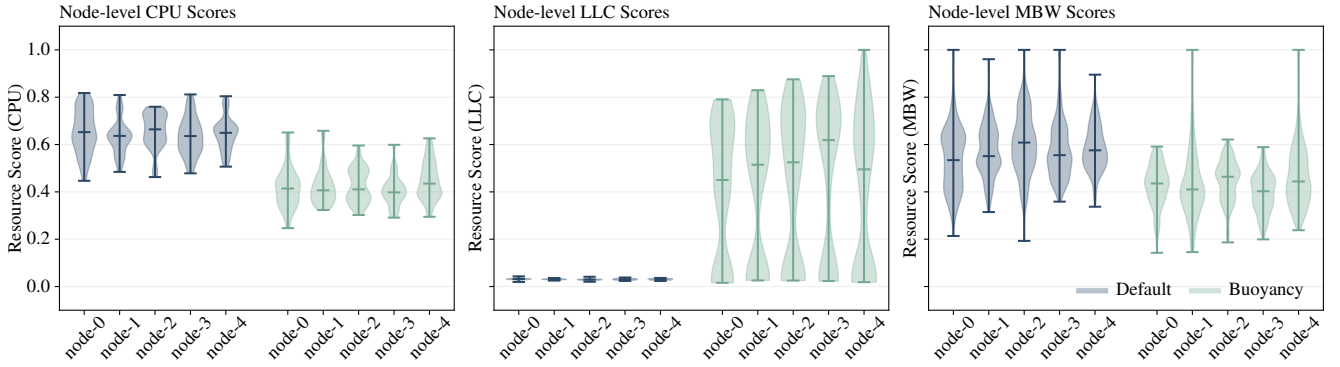
    }
    \caption{Resource scores for each worker node in the cluster for CPU, LLC, and MBW across the two configurations. 
        The resource score resembles the utilization of a resource, where a score of $1$ indicates that the resource is 
        fully utilized and a score of $0$ indicates that the resource is completely idle.}
    \label{fig:resource-scores}
\end{figure*}

This is further supported by the resource scores presented in \figurename~\ref{fig:resource-scores},
where individual resource scores for CPU, LLC, and MBW are shown for each worker node in the cluster.
For both CPU and MBW, our solution displays lower resource scores than the default configuration, 
which indicates that these resources experience less congestion. The difference being the LLC score which is
much higher in our solution than in the default configuration. This is due to the fact that the LLC resource
score gives a better indication of the actual resource pressure if workloads are separated into their own COSs,
which only happens for a workload that is classified as at risk in our solution, and never happens in the default
configuration. With this in mind, we can conclude that our solution is able to reduce resource 
fragmentation across the cluster and improve the distribution of workloads across nodes, which answers question 3.

\subsubsection{Rescheduling Overheads}

\begin{figure}[tbp]
    \centering
    \scalebox{.5}{
        \fontsize{14}{14}\selectfont 
        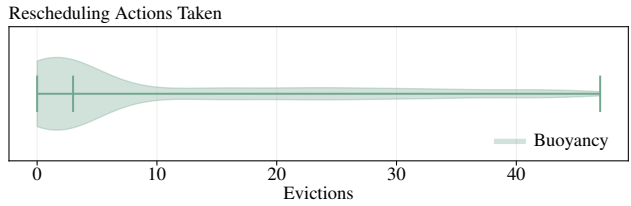
    }
    \caption{Total number of rescheduling actions taken by our controller at the time of measurement.
        As there is no rescheduling in the default configuration, 
        the number of rescheduling actions is $0$.
    }
    \label{fig:evictions}
\end{figure}

Finally, we examine the rescheduling overheads introduced by our solution.
By observing the number of rescheduling actions taken by our solution, we can gain an understanding of the overheads
required to maintain the performance of workloads in the cluster. The results are presented in \figurename~\ref{fig:evictions}
and display the total number of evictions performed by our solution from the start from the experiment at the time of measurement. 
As there is no rescheduling in the default configuration, the number of rescheduling actions is $0$.

We see that the number of rescheduling actions taken by our solution is relatively low. 
The median number of rescheduling actions is $3.00$, however, some iterations show a much higher number of rescheduling actions. 
This is to be expected due to the fact that workloads are randomly selected for deployment and some configurations will be less 
compatible with the current resource pressures on the nodes than others, and brings the average number of rescheduling actions to $11.0$ across all iterations.
This higher number of rescheduling actions does indicate a higher overhead, however, they are still 
infrequent enough in comparison to the total number of workloads deployed in the cluster
to be considered low.
Furthermore, the variation in rescheduling actions indicate that our solution 
is able to adapt to the current workload composition 
and resource pressures on the nodes, even as the workload dynamics evolve. 
Thus, in answer to question 4, we conclude that the 
rescheduling overheads introduced by our solution are low and do not greatly impact the performance of workloads in the cluster.

\input{content/75.limitations.tex}

%% file: slo-buoyancy-g5k_svg-tex.pdf_tex
\begingroup%
  \makeatletter%
  \providecommand\color[2][]{%
    \errmessage{(Inkscape) Color is used for the text in Inkscape, but the package 'color.sty' is not loaded}%
    \renewcommand\color[2][]{}%
  }%
  \providecommand\transparent[1]{%
    \errmessage{(Inkscape) Transparency is used (non-zero) for the text in Inkscape, but the package 'transparent.sty' is not loaded}%
    \renewcommand\transparent[1]{}%
  }%
  \providecommand\rotatebox[2]{#2}%
  \newcommand*\fsize{\dimexpr\f@size pt\relax}%
  \newcommand*\lineheight[1]{\fontsize{\fsize}{#1\fsize}\selectfont}%
  \ifx\svgwidth\undefined%
    \setlength{\unitlength}{487.84020996bp}%
    \ifx\svgscale\undefined%
      \relax%
    \else%
      \setlength{\unitlength}{\unitlength * \real{\svgscale}}%
    \fi%
  \else%
    \setlength{\unitlength}{\svgwidth}%
  \fi%
  \global\let\svgwidth\undefined%
  \global\let\svgscale\undefined%
  \makeatother%
  \begin{picture}(1,1.02860784)%
    \lineheight{1}%
    \setlength\tabcolsep{0pt}%
    \put(0,0){\includegraphics[width=\unitlength,page=1]{slo-buoyancy-g5k_svg-tex.pdf}}%
    \put(0.19588859,0.72105132){\makebox(0,0)[t]{\lineheight{0}\smash{\begin{tabular}[t]{c}-40000\end{tabular}}}}%
    \put(0,0){\includegraphics[width=\unitlength,page=2]{slo-buoyancy-g5k_svg-tex.pdf}}%
    \put(0.37763344,0.72105132){\makebox(0,0)[t]{\lineheight{0}\smash{\begin{tabular}[t]{c}-30000\end{tabular}}}}%
    \put(0,0){\includegraphics[width=\unitlength,page=3]{slo-buoyancy-g5k_svg-tex.pdf}}%
    \put(0.55937826,0.72105132){\makebox(0,0)[t]{\lineheight{0}\smash{\begin{tabular}[t]{c}-20000\end{tabular}}}}%
    \put(0,0){\includegraphics[width=\unitlength,page=4]{slo-buoyancy-g5k_svg-tex.pdf}}%
    \put(0.74112311,0.72105132){\makebox(0,0)[t]{\lineheight{0}\smash{\begin{tabular}[t]{c}-10000\end{tabular}}}}%
    \put(0,0){\includegraphics[width=\unitlength,page=5]{slo-buoyancy-g5k_svg-tex.pdf}}%
    \put(0.92286796,0.72105132){\makebox(0,0)[t]{\lineheight{0}\smash{\begin{tabular}[t]{c}0\end{tabular}}}}%
    \put(0.50402836,0.69301318){\makebox(0,0)[t]{\lineheight{0}\smash{\begin{tabular}[t]{c}SLO Slack (\%)\end{tabular}}}}%
    \put(0,0){\includegraphics[width=\unitlength,page=6]{slo-buoyancy-g5k_svg-tex.pdf}}%
    \put(0.04180753,0.98424905){\makebox(0,0)[lt]{\lineheight{0}\smash{\begin{tabular}[t]{l}SLO Slack\end{tabular}}}}%
    \put(0,0){\includegraphics[width=\unitlength,page=7]{slo-buoyancy-g5k_svg-tex.pdf}}%
    \put(0.05671788,0.38405571){\makebox(0,0)[t]{\lineheight{0}\smash{\begin{tabular}[t]{c}-500\end{tabular}}}}%
    \put(0,0){\includegraphics[width=\unitlength,page=8]{slo-buoyancy-g5k_svg-tex.pdf}}%
    \put(0.20582138,0.38405571){\makebox(0,0)[t]{\lineheight{0}\smash{\begin{tabular}[t]{c}-400\end{tabular}}}}%
    \put(0,0){\includegraphics[width=\unitlength,page=9]{slo-buoyancy-g5k_svg-tex.pdf}}%
    \put(0.35492487,0.38405571){\makebox(0,0)[t]{\lineheight{0}\smash{\begin{tabular}[t]{c}-300\end{tabular}}}}%
    \put(0,0){\includegraphics[width=\unitlength,page=10]{slo-buoyancy-g5k_svg-tex.pdf}}%
    \put(0.50402836,0.38405571){\makebox(0,0)[t]{\lineheight{0}\smash{\begin{tabular}[t]{c}-200\end{tabular}}}}%
    \put(0,0){\includegraphics[width=\unitlength,page=11]{slo-buoyancy-g5k_svg-tex.pdf}}%
    \put(0.65313188,0.38405571){\makebox(0,0)[t]{\lineheight{0}\smash{\begin{tabular}[t]{c}-100\end{tabular}}}}%
    \put(0,0){\includegraphics[width=\unitlength,page=12]{slo-buoyancy-g5k_svg-tex.pdf}}%
    \put(0.80223537,0.38405571){\makebox(0,0)[t]{\lineheight{0}\smash{\begin{tabular}[t]{c}0\end{tabular}}}}%
    \put(0,0){\includegraphics[width=\unitlength,page=13]{slo-buoyancy-g5k_svg-tex.pdf}}%
    \put(0.95133886,0.38405571){\makebox(0,0)[t]{\lineheight{0}\smash{\begin{tabular}[t]{c}100\end{tabular}}}}%
    \put(0.50402836,0.35601757){\makebox(0,0)[t]{\lineheight{0}\smash{\begin{tabular}[t]{c}SLO Slack (\%)\end{tabular}}}}%
    \put(0,0){\includegraphics[width=\unitlength,page=14]{slo-buoyancy-g5k_svg-tex.pdf}}%
    \put(0.04180753,0.64725348){\makebox(0,0)[lt]{\lineheight{0}\smash{\begin{tabular}[t]{l}SLO Slack (Scaled)\end{tabular}}}}%
    \put(0,0){\includegraphics[width=\unitlength,page=15]{slo-buoyancy-g5k_svg-tex.pdf}}%
    \put(0.0461681,0.04706013){\makebox(0,0)[t]{\lineheight{0}\smash{\begin{tabular}[t]{c}−20.0\end{tabular}}}}%
    \put(0,0){\includegraphics[width=\unitlength,page=16]{slo-buoyancy-g5k_svg-tex.pdf}}%
    \put(0.15518245,0.04706013){\makebox(0,0)[t]{\lineheight{0}\smash{\begin{tabular}[t]{c}−17.5\end{tabular}}}}%
    \put(0,0){\includegraphics[width=\unitlength,page=17]{slo-buoyancy-g5k_svg-tex.pdf}}%
    \put(0.26419679,0.04706013){\makebox(0,0)[t]{\lineheight{0}\smash{\begin{tabular}[t]{c}−15.0\end{tabular}}}}%
    \put(0,0){\includegraphics[width=\unitlength,page=18]{slo-buoyancy-g5k_svg-tex.pdf}}%
    \put(0.37321113,0.04706013){\makebox(0,0)[t]{\lineheight{0}\smash{\begin{tabular}[t]{c}−12.5\end{tabular}}}}%
    \put(0,0){\includegraphics[width=\unitlength,page=19]{slo-buoyancy-g5k_svg-tex.pdf}}%
    \put(0.48222548,0.04706013){\makebox(0,0)[t]{\lineheight{0}\smash{\begin{tabular}[t]{c}−10.0\end{tabular}}}}%
    \put(0,0){\includegraphics[width=\unitlength,page=20]{slo-buoyancy-g5k_svg-tex.pdf}}%
    \put(0.59123985,0.04706013){\makebox(0,0)[t]{\lineheight{0}\smash{\begin{tabular}[t]{c}−7.5\end{tabular}}}}%
    \put(0,0){\includegraphics[width=\unitlength,page=21]{slo-buoyancy-g5k_svg-tex.pdf}}%
    \put(0.70025419,0.04706013){\makebox(0,0)[t]{\lineheight{0}\smash{\begin{tabular}[t]{c}−5.0\end{tabular}}}}%
    \put(0,0){\includegraphics[width=\unitlength,page=22]{slo-buoyancy-g5k_svg-tex.pdf}}%
    \put(0.80926853,0.04706013){\makebox(0,0)[t]{\lineheight{0}\smash{\begin{tabular}[t]{c}−2.5\end{tabular}}}}%
    \put(0,0){\includegraphics[width=\unitlength,page=23]{slo-buoyancy-g5k_svg-tex.pdf}}%
    \put(0.91828288,0.04706013){\makebox(0,0)[t]{\lineheight{0}\smash{\begin{tabular}[t]{c}0.0\end{tabular}}}}%
    \put(0.50402836,0.01902199){\makebox(0,0)[t]{\lineheight{0}\smash{\begin{tabular}[t]{c}Buoyancy Score\end{tabular}}}}%
    \put(0,0){\includegraphics[width=\unitlength,page=24]{slo-buoyancy-g5k_svg-tex.pdf}}%
    \put(0.04180753,0.31025787){\makebox(0,0)[lt]{\lineheight{0}\smash{\begin{tabular}[t]{l}Buoyancy Score\end{tabular}}}}%
    \put(0,0){\includegraphics[width=\unitlength,page=25]{slo-buoyancy-g5k_svg-tex.pdf}}%
    \put(0.68151714,0.99458351){\makebox(0,0)[lt]{\lineheight{0}\smash{\begin{tabular}[t]{l}Default\end{tabular}}}}%
    \put(0,0){\includegraphics[width=\unitlength,page=26]{slo-buoyancy-g5k_svg-tex.pdf}}%
    \put(0.85480642,0.99458351){\makebox(0,0)[lt]{\lineheight{0}\smash{\begin{tabular}[t]{l}Buoyancy\end{tabular}}}}%
  \end{picture}%
\endgroup%

%% file: allocated-compliance-g5k_svg-tex.pdf_tex
\begingroup%
  \makeatletter%
  \providecommand\color[2][]{%
    \errmessage{(Inkscape) Color is used for the text in Inkscape, but the package 'color.sty' is not loaded}%
    \renewcommand\color[2][]{}%
  }%
  \providecommand\transparent[1]{%
    \errmessage{(Inkscape) Transparency is used (non-zero) for the text in Inkscape, but the package 'transparent.sty' is not loaded}%
    \renewcommand\transparent[1]{}%
  }%
  \providecommand\rotatebox[2]{#2}%
  \newcommand*\fsize{\dimexpr\f@size pt\relax}%
  \newcommand*\lineheight[1]{\fontsize{\fsize}{#1\fsize}\selectfont}%
  \ifx\svgwidth\undefined%
    \setlength{\unitlength}{488.20001221bp}%
    \ifx\svgscale\undefined%
      \relax%
    \else%
      \setlength{\unitlength}{\unitlength * \real{\svgscale}}%
    \fi%
  \else%
    \setlength{\unitlength}{\svgwidth}%
  \fi%
  \global\let\svgwidth\undefined%
  \global\let\svgscale\undefined%
  \makeatother%
  \begin{picture}(1,0.68372851)%
    \lineheight{1}%
    \setlength\tabcolsep{0pt}%
    \put(0,0){\includegraphics[width=\unitlength,page=1]{allocated-compliance-g5k_svg-tex.pdf}}%
    \put(0.0656299,0.38008567){\makebox(0,0)[t]{\lineheight{0}\smash{\begin{tabular}[t]{c}70\end{tabular}}}}%
    \put(0,0){\includegraphics[width=\unitlength,page=2]{allocated-compliance-g5k_svg-tex.pdf}}%
    \put(0.36083222,0.38008567){\makebox(0,0)[t]{\lineheight{0}\smash{\begin{tabular}[t]{c}80\end{tabular}}}}%
    \put(0,0){\includegraphics[width=\unitlength,page=3]{allocated-compliance-g5k_svg-tex.pdf}}%
    \put(0.65603452,0.38008567){\makebox(0,0)[t]{\lineheight{0}\smash{\begin{tabular}[t]{c}90\end{tabular}}}}%
    \put(0,0){\includegraphics[width=\unitlength,page=4]{allocated-compliance-g5k_svg-tex.pdf}}%
    \put(0.95123682,0.38008567){\makebox(0,0)[t]{\lineheight{0}\smash{\begin{tabular}[t]{c}100\end{tabular}}}}%
    \put(0.49367326,0.3520682){\makebox(0,0)[t]{\lineheight{0}\smash{\begin{tabular}[t]{c}CPU Cores Allocated (\%)\end{tabular}}}}%
    \put(0,0){\includegraphics[width=\unitlength,page=5]{allocated-compliance-g5k_svg-tex.pdf}}%
    \put(0.02191725,0.63940241){\makebox(0,0)[lt]{\lineheight{0}\smash{\begin{tabular}[t]{l}Allocated CPU\end{tabular}}}}%
    \put(0,0){\includegraphics[width=\unitlength,page=6]{allocated-compliance-g5k_svg-tex.pdf}}%
    \put(0.06480416,0.04702545){\makebox(0,0)[t]{\lineheight{0}\smash{\begin{tabular}[t]{c}0\end{tabular}}}}%
    \put(0,0){\includegraphics[width=\unitlength,page=7]{allocated-compliance-g5k_svg-tex.pdf}}%
    \put(0.23635179,0.04702545){\makebox(0,0)[t]{\lineheight{0}\smash{\begin{tabular}[t]{c}20\end{tabular}}}}%
    \put(0,0){\includegraphics[width=\unitlength,page=8]{allocated-compliance-g5k_svg-tex.pdf}}%
    \put(0.40789943,0.04702545){\makebox(0,0)[t]{\lineheight{0}\smash{\begin{tabular}[t]{c}40\end{tabular}}}}%
    \put(0,0){\includegraphics[width=\unitlength,page=9]{allocated-compliance-g5k_svg-tex.pdf}}%
    \put(0.57944709,0.04702545){\makebox(0,0)[t]{\lineheight{0}\smash{\begin{tabular}[t]{c}60\end{tabular}}}}%
    \put(0,0){\includegraphics[width=\unitlength,page=10]{allocated-compliance-g5k_svg-tex.pdf}}%
    \put(0.7509947,0.04702545){\makebox(0,0)[t]{\lineheight{0}\smash{\begin{tabular}[t]{c}80\end{tabular}}}}%
    \put(0,0){\includegraphics[width=\unitlength,page=11]{allocated-compliance-g5k_svg-tex.pdf}}%
    \put(0.92254236,0.04702545){\makebox(0,0)[t]{\lineheight{0}\smash{\begin{tabular}[t]{c}100\end{tabular}}}}%
    \put(0.49367326,0.01900797){\makebox(0,0)[t]{\lineheight{0}\smash{\begin{tabular}[t]{c}Allocated CPU Cores Within SLO (\%)\end{tabular}}}}%
    \put(0,0){\includegraphics[width=\unitlength,page=12]{allocated-compliance-g5k_svg-tex.pdf}}%
    \put(0.02191725,0.30634219){\makebox(0,0)[lt]{\lineheight{0}\smash{\begin{tabular}[t]{l}CPU Within SLO\end{tabular}}}}%
    \put(0,0){\includegraphics[width=\unitlength,page=13]{allocated-compliance-g5k_svg-tex.pdf}}%
    \put(0.6817518,0.64972925){\makebox(0,0)[lt]{\lineheight{0}\smash{\begin{tabular}[t]{l}Default\end{tabular}}}}%
    \put(0,0){\includegraphics[width=\unitlength,page=14]{allocated-compliance-g5k_svg-tex.pdf}}%
    \put(0.85491342,0.64972925){\makebox(0,0)[lt]{\lineheight{0}\smash{\begin{tabular}[t]{l}Buoyancy\end{tabular}}}}%
    \put(0,0){\includegraphics[width=\unitlength,page=15]{allocated-compliance-g5k_svg-tex.pdf}}%
    \put(0.6817518,0.64972925){\makebox(0,0)[lt]{\lineheight{0}\smash{\begin{tabular}[t]{l}Default\end{tabular}}}}%
    \put(0,0){\includegraphics[width=\unitlength,page=16]{allocated-compliance-g5k_svg-tex.pdf}}%
    \put(0.85491342,0.64972925){\makebox(0,0)[lt]{\lineheight{0}\smash{\begin{tabular}[t]{l}Buoyancy\end{tabular}}}}%
  \end{picture}%
\endgroup%

%% file: node-buoyancy-g5k_svg-tex.pdf_tex
\begingroup%
  \makeatletter%
  \providecommand\color[2][]{%
    \errmessage{(Inkscape) Color is used for the text in Inkscape, but the package 'color.sty' is not loaded}%
    \renewcommand\color[2][]{}%
  }%
  \providecommand\transparent[1]{%
    \errmessage{(Inkscape) Transparency is used (non-zero) for the text in Inkscape, but the package 'transparent.sty' is not loaded}%
    \renewcommand\transparent[1]{}%
  }%
  \providecommand\rotatebox[2]{#2}%
  \newcommand*\fsize{\dimexpr\f@size pt\relax}%
  \newcommand*\lineheight[1]{\fontsize{\fsize}{#1\fsize}\selectfont}%
  \ifx\svgwidth\undefined%
    \setlength{\unitlength}{1000.32531738bp}%
    \ifx\svgscale\undefined%
      \relax%
    \else%
      \setlength{\unitlength}{\unitlength * \real{\svgscale}}%
    \fi%
  \else%
    \setlength{\unitlength}{\svgwidth}%
  \fi%
  \global\let\svgwidth\undefined%
  \global\let\svgscale\undefined%
  \makeatother%
  \begin{picture}(1,0.28078491)%
    \lineheight{1}%
    \setlength\tabcolsep{0pt}%
    \put(0,0){\includegraphics[width=\unitlength,page=1]{node-buoyancy-g5k_svg-tex.pdf}}%
    \put(0.09338866,0.00927668){\makebox(0,0)[t]{\lineheight{0}\smash{\begin{tabular}[t]{c}node-0\end{tabular}}}}%
    \put(0,0){\includegraphics[width=\unitlength,page=2]{node-buoyancy-g5k_svg-tex.pdf}}%
    \put(0.14776199,0.00927668){\makebox(0,0)[t]{\lineheight{0}\smash{\begin{tabular}[t]{c}node-1\end{tabular}}}}%
    \put(0,0){\includegraphics[width=\unitlength,page=3]{node-buoyancy-g5k_svg-tex.pdf}}%
    \put(0.20213534,0.00927668){\makebox(0,0)[t]{\lineheight{0}\smash{\begin{tabular}[t]{c}node-2\end{tabular}}}}%
    \put(0,0){\includegraphics[width=\unitlength,page=4]{node-buoyancy-g5k_svg-tex.pdf}}%
    \put(0.25650869,0.00927668){\makebox(0,0)[t]{\lineheight{0}\smash{\begin{tabular}[t]{c}node-3\end{tabular}}}}%
    \put(0,0){\includegraphics[width=\unitlength,page=5]{node-buoyancy-g5k_svg-tex.pdf}}%
    \put(0.31088202,0.00927668){\makebox(0,0)[t]{\lineheight{0}\smash{\begin{tabular}[t]{c}node-4\end{tabular}}}}%
    \put(0,0){\includegraphics[width=\unitlength,page=6]{node-buoyancy-g5k_svg-tex.pdf}}%
    \put(0.40603536,0.00927668){\makebox(0,0)[t]{\lineheight{0}\smash{\begin{tabular}[t]{c}node-0\end{tabular}}}}%
    \put(0,0){\includegraphics[width=\unitlength,page=7]{node-buoyancy-g5k_svg-tex.pdf}}%
    \put(0.46040872,0.00927668){\makebox(0,0)[t]{\lineheight{0}\smash{\begin{tabular}[t]{c}node-1\end{tabular}}}}%
    \put(0,0){\includegraphics[width=\unitlength,page=8]{node-buoyancy-g5k_svg-tex.pdf}}%
    \put(0.51478206,0.00927668){\makebox(0,0)[t]{\lineheight{0}\smash{\begin{tabular}[t]{c}node-2\end{tabular}}}}%
    \put(0,0){\includegraphics[width=\unitlength,page=9]{node-buoyancy-g5k_svg-tex.pdf}}%
    \put(0.56915542,0.00927668){\makebox(0,0)[t]{\lineheight{0}\smash{\begin{tabular}[t]{c}node-3\end{tabular}}}}%
    \put(0,0){\includegraphics[width=\unitlength,page=10]{node-buoyancy-g5k_svg-tex.pdf}}%
    \put(0.62352872,0.00927668){\makebox(0,0)[t]{\lineheight{0}\smash{\begin{tabular}[t]{c}node-4\end{tabular}}}}%
    \put(0,0){\includegraphics[width=\unitlength,page=11]{node-buoyancy-g5k_svg-tex.pdf}}%
    \put(0.04832959,0.04057083){\makebox(0,0)[rt]{\lineheight{0}\smash{\begin{tabular}[t]{r}−175\end{tabular}}}}%
    \put(0,0){\includegraphics[width=\unitlength,page=12]{node-buoyancy-g5k_svg-tex.pdf}}%
    \put(0.04832959,0.0695969){\makebox(0,0)[rt]{\lineheight{0}\smash{\begin{tabular}[t]{r}−150\end{tabular}}}}%
    \put(0,0){\includegraphics[width=\unitlength,page=13]{node-buoyancy-g5k_svg-tex.pdf}}%
    \put(0.04832959,0.09862296){\makebox(0,0)[rt]{\lineheight{0}\smash{\begin{tabular}[t]{r}−125\end{tabular}}}}%
    \put(0,0){\includegraphics[width=\unitlength,page=14]{node-buoyancy-g5k_svg-tex.pdf}}%
    \put(0.04832959,0.12764903){\makebox(0,0)[rt]{\lineheight{0}\smash{\begin{tabular}[t]{r}−100\end{tabular}}}}%
    \put(0,0){\includegraphics[width=\unitlength,page=15]{node-buoyancy-g5k_svg-tex.pdf}}%
    \put(0.04832959,0.15667508){\makebox(0,0)[rt]{\lineheight{0}\smash{\begin{tabular}[t]{r}−75\end{tabular}}}}%
    \put(0,0){\includegraphics[width=\unitlength,page=16]{node-buoyancy-g5k_svg-tex.pdf}}%
    \put(0.04832959,0.18570115){\makebox(0,0)[rt]{\lineheight{0}\smash{\begin{tabular}[t]{r}−50\end{tabular}}}}%
    \put(0,0){\includegraphics[width=\unitlength,page=17]{node-buoyancy-g5k_svg-tex.pdf}}%
    \put(0.04832959,0.21472721){\makebox(0,0)[rt]{\lineheight{0}\smash{\begin{tabular}[t]{r}−25\end{tabular}}}}%
    \put(0,0){\includegraphics[width=\unitlength,page=18]{node-buoyancy-g5k_svg-tex.pdf}}%
    \put(0.04832959,0.24375327){\makebox(0,0)[rt]{\lineheight{0}\smash{\begin{tabular}[t]{r}0\end{tabular}}}}%
    \put(0.01479363,0.14117221){\rotatebox{90}{\makebox(0,0)[t]{\lineheight{0}\smash{\begin{tabular}[t]{c}Node Buoyancy Score\end{tabular}}}}}%
    \put(0,0){\includegraphics[width=\unitlength,page=19]{node-buoyancy-g5k_svg-tex.pdf}}%
    \put(0.05532732,0.26447209){\makebox(0,0)[lt]{\lineheight{0}\smash{\begin{tabular}[t]{l}Node-level Buoyancy Scores\end{tabular}}}}%
    \put(0,0){\includegraphics[width=\unitlength,page=20]{node-buoyancy-g5k_svg-tex.pdf}}%
    \put(0.50627185,0.03494644){\makebox(0,0)[lt]{\lineheight{0}\smash{\begin{tabular}[t]{l}Default\end{tabular}}}}%
    \put(0,0){\includegraphics[width=\unitlength,page=21]{node-buoyancy-g5k_svg-tex.pdf}}%
    \put(0.59078187,0.03494644){\makebox(0,0)[lt]{\lineheight{0}\smash{\begin{tabular}[t]{l}Buoyancy\end{tabular}}}}%
    \put(0,0){\includegraphics[width=\unitlength,page=22]{node-buoyancy-g5k_svg-tex.pdf}}%
    \put(0.74887454,0.00927668){\makebox(0,0)[t]{\lineheight{0}\smash{\begin{tabular}[t]{c}node-0\end{tabular}}}}%
    \put(0,0){\includegraphics[width=\unitlength,page=23]{node-buoyancy-g5k_svg-tex.pdf}}%
    \put(0.8020758,0.00927668){\makebox(0,0)[t]{\lineheight{0}\smash{\begin{tabular}[t]{c}node-1\end{tabular}}}}%
    \put(0,0){\includegraphics[width=\unitlength,page=24]{node-buoyancy-g5k_svg-tex.pdf}}%
    \put(0.85527706,0.00927668){\makebox(0,0)[t]{\lineheight{0}\smash{\begin{tabular}[t]{c}node-2\end{tabular}}}}%
    \put(0,0){\includegraphics[width=\unitlength,page=25]{node-buoyancy-g5k_svg-tex.pdf}}%
    \put(0.90847832,0.00927668){\makebox(0,0)[t]{\lineheight{0}\smash{\begin{tabular}[t]{c}node-3\end{tabular}}}}%
    \put(0,0){\includegraphics[width=\unitlength,page=26]{node-buoyancy-g5k_svg-tex.pdf}}%
    \put(0.96167958,0.00927668){\makebox(0,0)[t]{\lineheight{0}\smash{\begin{tabular}[t]{c}node-4\end{tabular}}}}%
    \put(0,0){\includegraphics[width=\unitlength,page=27]{node-buoyancy-g5k_svg-tex.pdf}}%
    \put(0.71075407,0.0287614){\makebox(0,0)[rt]{\lineheight{0}\smash{\begin{tabular}[t]{r}−2.0\end{tabular}}}}%
    \put(0,0){\includegraphics[width=\unitlength,page=28]{node-buoyancy-g5k_svg-tex.pdf}}%
    \put(0.71075407,0.07220652){\makebox(0,0)[rt]{\lineheight{0}\smash{\begin{tabular}[t]{r}−1.5\end{tabular}}}}%
    \put(0,0){\includegraphics[width=\unitlength,page=29]{node-buoyancy-g5k_svg-tex.pdf}}%
    \put(0.71075407,0.11565166){\makebox(0,0)[rt]{\lineheight{0}\smash{\begin{tabular}[t]{r}−1.0\end{tabular}}}}%
    \put(0,0){\includegraphics[width=\unitlength,page=30]{node-buoyancy-g5k_svg-tex.pdf}}%
    \put(0.71075407,0.15909679){\makebox(0,0)[rt]{\lineheight{0}\smash{\begin{tabular}[t]{r}−0.5\end{tabular}}}}%
    \put(0,0){\includegraphics[width=\unitlength,page=31]{node-buoyancy-g5k_svg-tex.pdf}}%
    \put(0.71075407,0.20254191){\makebox(0,0)[rt]{\lineheight{0}\smash{\begin{tabular}[t]{r}0.0\end{tabular}}}}%
    \put(0,0){\includegraphics[width=\unitlength,page=32]{node-buoyancy-g5k_svg-tex.pdf}}%
    \put(0.71075407,0.24598703){\makebox(0,0)[rt]{\lineheight{0}\smash{\begin{tabular}[t]{r}0.5\end{tabular}}}}%
    \put(0.68040146,0.14117219){\rotatebox{90}{\makebox(0,0)[t]{\lineheight{0}\smash{\begin{tabular}[t]{c}Node Buoyancy Score\end{tabular}}}}}%
    \put(0,0){\includegraphics[width=\unitlength,page=33]{node-buoyancy-g5k_svg-tex.pdf}}%
    \put(0.71775179,0.26447209){\makebox(0,0)[lt]{\lineheight{0}\smash{\begin{tabular}[t]{l}Node-level Buoyancy Scores (Scaled)\end{tabular}}}}%
  \end{picture}%
\endgroup%

%% file: resource-scores-g5k_svg-tex.pdf_tex
\begingroup%
  \makeatletter%
  \providecommand\color[2][]{%
    \errmessage{(Inkscape) Color is used for the text in Inkscape, but the package 'color.sty' is not loaded}%
    \renewcommand\color[2][]{}%
  }%
  \providecommand\transparent[1]{%
    \errmessage{(Inkscape) Transparency is used (non-zero) for the text in Inkscape, but the package 'transparent.sty' is not loaded}%
    \renewcommand\transparent[1]{}%
  }%
  \providecommand\rotatebox[2]{#2}%
  \newcommand*\fsize{\dimexpr\f@size pt\relax}%
  \newcommand*\lineheight[1]{\fontsize{\fsize}{#1\fsize}\selectfont}%
  \ifx\svgwidth\undefined%
    \setlength{\unitlength}{998.91784668bp}%
    \ifx\svgscale\undefined%
      \relax%
    \else%
      \setlength{\unitlength}{\unitlength * \real{\svgscale}}%
    \fi%
  \else%
    \setlength{\unitlength}{\svgwidth}%
  \fi%
  \global\let\svgwidth\undefined%
  \global\let\svgscale\undefined%
  \makeatother%
  \begin{picture}(1,0.28136784)%
    \lineheight{1}%
    \setlength\tabcolsep{0pt}%
    \put(0,0){\includegraphics[width=\unitlength,page=1]{resource-scores-g5k_svg-tex.pdf}}%
    \put(0.05202838,0.00867997){\rotatebox{45}{\makebox(0,0)[lt]{\lineheight{0}\smash{\begin{tabular}[t]{l}node-0\end{tabular}}}}}%
    \put(0,0){\includegraphics[width=\unitlength,page=2]{resource-scores-g5k_svg-tex.pdf}}%
    \put(0.07861229,0.00867997){\rotatebox{45}{\makebox(0,0)[lt]{\lineheight{0}\smash{\begin{tabular}[t]{l}node-1\end{tabular}}}}}%
    \put(0,0){\includegraphics[width=\unitlength,page=3]{resource-scores-g5k_svg-tex.pdf}}%
    \put(0.10519621,0.00867997){\rotatebox{45}{\makebox(0,0)[lt]{\lineheight{0}\smash{\begin{tabular}[t]{l}node-2\end{tabular}}}}}%
    \put(0,0){\includegraphics[width=\unitlength,page=4]{resource-scores-g5k_svg-tex.pdf}}%
    \put(0.13178014,0.00867996){\rotatebox{45}{\makebox(0,0)[lt]{\lineheight{0}\smash{\begin{tabular}[t]{l}node-3\end{tabular}}}}}%
    \put(0,0){\includegraphics[width=\unitlength,page=5]{resource-scores-g5k_svg-tex.pdf}}%
    \put(0.15836405,0.00867997){\rotatebox{45}{\makebox(0,0)[lt]{\lineheight{0}\smash{\begin{tabular}[t]{l}node-4\end{tabular}}}}}%
    \put(0,0){\includegraphics[width=\unitlength,page=6]{resource-scores-g5k_svg-tex.pdf}}%
    \put(0.20488591,0.00867997){\rotatebox{45}{\makebox(0,0)[lt]{\lineheight{0}\smash{\begin{tabular}[t]{l}node-0\end{tabular}}}}}%
    \put(0,0){\includegraphics[width=\unitlength,page=7]{resource-scores-g5k_svg-tex.pdf}}%
    \put(0.23146983,0.00867997){\rotatebox{45}{\makebox(0,0)[lt]{\lineheight{0}\smash{\begin{tabular}[t]{l}node-1\end{tabular}}}}}%
    \put(0,0){\includegraphics[width=\unitlength,page=8]{resource-scores-g5k_svg-tex.pdf}}%
    \put(0.25805375,0.00867997){\rotatebox{45}{\makebox(0,0)[lt]{\lineheight{0}\smash{\begin{tabular}[t]{l}node-2\end{tabular}}}}}%
    \put(0,0){\includegraphics[width=\unitlength,page=9]{resource-scores-g5k_svg-tex.pdf}}%
    \put(0.28463767,0.00867997){\rotatebox{45}{\makebox(0,0)[lt]{\lineheight{0}\smash{\begin{tabular}[t]{l}node-3\end{tabular}}}}}%
    \put(0,0){\includegraphics[width=\unitlength,page=10]{resource-scores-g5k_svg-tex.pdf}}%
    \put(0.31122159,0.00867997){\rotatebox{45}{\makebox(0,0)[lt]{\lineheight{0}\smash{\begin{tabular}[t]{l}node-4\end{tabular}}}}}%
    \put(0,0){\includegraphics[width=\unitlength,page=11]{resource-scores-g5k_svg-tex.pdf}}%
    \put(0.0368211,0.05975716){\makebox(0,0)[rt]{\lineheight{0}\smash{\begin{tabular}[t]{r}0.0\end{tabular}}}}%
    \put(0,0){\includegraphics[width=\unitlength,page=12]{resource-scores-g5k_svg-tex.pdf}}%
    \put(0.0368211,0.09529626){\makebox(0,0)[rt]{\lineheight{0}\smash{\begin{tabular}[t]{r}0.2\end{tabular}}}}%
    \put(0,0){\includegraphics[width=\unitlength,page=13]{resource-scores-g5k_svg-tex.pdf}}%
    \put(0.0368211,0.13083536){\makebox(0,0)[rt]{\lineheight{0}\smash{\begin{tabular}[t]{r}0.4\end{tabular}}}}%
    \put(0,0){\includegraphics[width=\unitlength,page=14]{resource-scores-g5k_svg-tex.pdf}}%
    \put(0.0368211,0.16637445){\makebox(0,0)[rt]{\lineheight{0}\smash{\begin{tabular}[t]{r}0.6\end{tabular}}}}%
    \put(0,0){\includegraphics[width=\unitlength,page=15]{resource-scores-g5k_svg-tex.pdf}}%
    \put(0.0368211,0.20191355){\makebox(0,0)[rt]{\lineheight{0}\smash{\begin{tabular}[t]{r}0.8\end{tabular}}}}%
    \put(0,0){\includegraphics[width=\unitlength,page=16]{resource-scores-g5k_svg-tex.pdf}}%
    \put(0.0368211,0.23745265){\makebox(0,0)[rt]{\lineheight{0}\smash{\begin{tabular}[t]{r}1.0\end{tabular}}}}%
    \put(0.01481447,0.15240824){\rotatebox{90}{\makebox(0,0)[t]{\lineheight{0}\smash{\begin{tabular}[t]{c}Resource Score (CPU)\end{tabular}}}}}%
    \put(0,0){\includegraphics[width=\unitlength,page=17]{resource-scores-g5k_svg-tex.pdf}}%
    \put(0.04382868,0.26503204){\makebox(0,0)[lt]{\lineheight{0}\smash{\begin{tabular}[t]{l}Node-level CPU Scores\end{tabular}}}}%
    \put(0,0){\includegraphics[width=\unitlength,page=18]{resource-scores-g5k_svg-tex.pdf}}%
    \put(0.37830479,0.00867997){\rotatebox{45}{\makebox(0,0)[lt]{\lineheight{0}\smash{\begin{tabular}[t]{l}node-0\end{tabular}}}}}%
    \put(0,0){\includegraphics[width=\unitlength,page=19]{resource-scores-g5k_svg-tex.pdf}}%
    \put(0.40488871,0.00867997){\rotatebox{45}{\makebox(0,0)[lt]{\lineheight{0}\smash{\begin{tabular}[t]{l}node-1\end{tabular}}}}}%
    \put(0,0){\includegraphics[width=\unitlength,page=20]{resource-scores-g5k_svg-tex.pdf}}%
    \put(0.43147263,0.00867996){\rotatebox{45}{\makebox(0,0)[lt]{\lineheight{0}\smash{\begin{tabular}[t]{l}node-2\end{tabular}}}}}%
    \put(0,0){\includegraphics[width=\unitlength,page=21]{resource-scores-g5k_svg-tex.pdf}}%
    \put(0.45805655,0.00867997){\rotatebox{45}{\makebox(0,0)[lt]{\lineheight{0}\smash{\begin{tabular}[t]{l}node-3\end{tabular}}}}}%
    \put(0,0){\includegraphics[width=\unitlength,page=22]{resource-scores-g5k_svg-tex.pdf}}%
    \put(0.48464047,0.00867997){\rotatebox{45}{\makebox(0,0)[lt]{\lineheight{0}\smash{\begin{tabular}[t]{l}node-4\end{tabular}}}}}%
    \put(0,0){\includegraphics[width=\unitlength,page=23]{resource-scores-g5k_svg-tex.pdf}}%
    \put(0.53116233,0.00867996){\rotatebox{45}{\makebox(0,0)[lt]{\lineheight{0}\smash{\begin{tabular}[t]{l}node-0\end{tabular}}}}}%
    \put(0,0){\includegraphics[width=\unitlength,page=24]{resource-scores-g5k_svg-tex.pdf}}%
    \put(0.55774625,0.00867997){\rotatebox{45}{\makebox(0,0)[lt]{\lineheight{0}\smash{\begin{tabular}[t]{l}node-1\end{tabular}}}}}%
    \put(0,0){\includegraphics[width=\unitlength,page=25]{resource-scores-g5k_svg-tex.pdf}}%
    \put(0.58433017,0.00867997){\rotatebox{45}{\makebox(0,0)[lt]{\lineheight{0}\smash{\begin{tabular}[t]{l}node-2\end{tabular}}}}}%
    \put(0,0){\includegraphics[width=\unitlength,page=26]{resource-scores-g5k_svg-tex.pdf}}%
    \put(0.61091408,0.00867997){\rotatebox{45}{\makebox(0,0)[lt]{\lineheight{0}\smash{\begin{tabular}[t]{l}node-3\end{tabular}}}}}%
    \put(0,0){\includegraphics[width=\unitlength,page=27]{resource-scores-g5k_svg-tex.pdf}}%
    \put(0.63749801,0.00867997){\rotatebox{45}{\makebox(0,0)[lt]{\lineheight{0}\smash{\begin{tabular}[t]{l}node-4\end{tabular}}}}}%
    \put(0,0){\includegraphics[width=\unitlength,page=28]{resource-scores-g5k_svg-tex.pdf}}%
    \put(0.36051503,0.15240825){\rotatebox{90}{\makebox(0,0)[t]{\lineheight{0}\smash{\begin{tabular}[t]{c}Resource Score (LLC)\end{tabular}}}}}%
    \put(0,0){\includegraphics[width=\unitlength,page=29]{resource-scores-g5k_svg-tex.pdf}}%
    \put(0.3701051,0.26503204){\makebox(0,0)[lt]{\lineheight{0}\smash{\begin{tabular}[t]{l}Node-level LLC Scores\end{tabular}}}}%
    \put(0,0){\includegraphics[width=\unitlength,page=30]{resource-scores-g5k_svg-tex.pdf}}%
    \put(0.7045812,0.00867997){\rotatebox{45}{\makebox(0,0)[lt]{\lineheight{0}\smash{\begin{tabular}[t]{l}node-0\end{tabular}}}}}%
    \put(0,0){\includegraphics[width=\unitlength,page=31]{resource-scores-g5k_svg-tex.pdf}}%
    \put(0.73116513,0.00867997){\rotatebox{45}{\makebox(0,0)[lt]{\lineheight{0}\smash{\begin{tabular}[t]{l}node-1\end{tabular}}}}}%
    \put(0,0){\includegraphics[width=\unitlength,page=32]{resource-scores-g5k_svg-tex.pdf}}%
    \put(0.75774904,0.00867997){\rotatebox{45}{\makebox(0,0)[lt]{\lineheight{0}\smash{\begin{tabular}[t]{l}node-2\end{tabular}}}}}%
    \put(0,0){\includegraphics[width=\unitlength,page=33]{resource-scores-g5k_svg-tex.pdf}}%
    \put(0.78433296,0.00867997){\rotatebox{45}{\makebox(0,0)[lt]{\lineheight{0}\smash{\begin{tabular}[t]{l}node-3\end{tabular}}}}}%
    \put(0,0){\includegraphics[width=\unitlength,page=34]{resource-scores-g5k_svg-tex.pdf}}%
    \put(0.81091688,0.00867997){\rotatebox{45}{\makebox(0,0)[lt]{\lineheight{0}\smash{\begin{tabular}[t]{l}node-4\end{tabular}}}}}%
    \put(0,0){\includegraphics[width=\unitlength,page=35]{resource-scores-g5k_svg-tex.pdf}}%
    \put(0.85743874,0.00867997){\rotatebox{45}{\makebox(0,0)[lt]{\lineheight{0}\smash{\begin{tabular}[t]{l}node-0\end{tabular}}}}}%
    \put(0,0){\includegraphics[width=\unitlength,page=36]{resource-scores-g5k_svg-tex.pdf}}%
    \put(0.88402266,0.00867997){\rotatebox{45}{\makebox(0,0)[lt]{\lineheight{0}\smash{\begin{tabular}[t]{l}node-1\end{tabular}}}}}%
    \put(0,0){\includegraphics[width=\unitlength,page=37]{resource-scores-g5k_svg-tex.pdf}}%
    \put(0.91060658,0.00867997){\rotatebox{45}{\makebox(0,0)[lt]{\lineheight{0}\smash{\begin{tabular}[t]{l}node-2\end{tabular}}}}}%
    \put(0,0){\includegraphics[width=\unitlength,page=38]{resource-scores-g5k_svg-tex.pdf}}%
    \put(0.9371905,0.00867997){\rotatebox{45}{\makebox(0,0)[lt]{\lineheight{0}\smash{\begin{tabular}[t]{l}node-3\end{tabular}}}}}%
    \put(0,0){\includegraphics[width=\unitlength,page=39]{resource-scores-g5k_svg-tex.pdf}}%
    \put(0.96377443,0.00867997){\rotatebox{45}{\makebox(0,0)[lt]{\lineheight{0}\smash{\begin{tabular}[t]{l}node-4\end{tabular}}}}}%
    \put(0,0){\includegraphics[width=\unitlength,page=40]{resource-scores-g5k_svg-tex.pdf}}%
    \put(0.68679144,0.15240823){\rotatebox{90}{\makebox(0,0)[t]{\lineheight{0}\smash{\begin{tabular}[t]{c}Resource Score (MBW)\end{tabular}}}}}%
    \put(0,0){\includegraphics[width=\unitlength,page=41]{resource-scores-g5k_svg-tex.pdf}}%
    \put(0.69638152,0.26503204){\makebox(0,0)[lt]{\lineheight{0}\smash{\begin{tabular}[t]{l}Node-level MBW Scores\end{tabular}}}}%
    \put(0,0){\includegraphics[width=\unitlength,page=42]{resource-scores-g5k_svg-tex.pdf}}%
    \put(0.83725516,0.05688263){\makebox(0,0)[lt]{\lineheight{0}\smash{\begin{tabular}[t]{l}Default\end{tabular}}}}%
    \put(0,0){\includegraphics[width=\unitlength,page=43]{resource-scores-g5k_svg-tex.pdf}}%
    \put(0.92188422,0.05688263){\makebox(0,0)[lt]{\lineheight{0}\smash{\begin{tabular}[t]{l}Buoyancy\end{tabular}}}}%
  \end{picture}%
\endgroup%

%% file: evictions-g5k_svg-tex.pdf_tex
\begingroup%
  \makeatletter%
  \providecommand\color[2][]{%
    \errmessage{(Inkscape) Color is used for the text in Inkscape, but the package 'color.sty' is not loaded}%
    \renewcommand\color[2][]{}%
  }%
  \providecommand\transparent[1]{%
    \errmessage{(Inkscape) Transparency is used (non-zero) for the text in Inkscape, but the package 'transparent.sty' is not loaded}%
    \renewcommand\transparent[1]{}%
  }%
  \providecommand\rotatebox[2]{#2}%
  \newcommand*\fsize{\dimexpr\f@size pt\relax}%
  \newcommand*\lineheight[1]{\fontsize{\fsize}{#1\fsize}\selectfont}%
  \ifx\svgwidth\undefined%
    \setlength{\unitlength}{482.3999939bp}%
    \ifx\svgscale\undefined%
      \relax%
    \else%
      \setlength{\unitlength}{\unitlength * \real{\svgscale}}%
    \fi%
  \else%
    \setlength{\unitlength}{\svgwidth}%
  \fi%
  \global\let\svgwidth\undefined%
  \global\let\svgscale\undefined%
  \makeatother%
  \begin{picture}(1,0.33265834)%
    \lineheight{1}%
    \setlength\tabcolsep{0pt}%
    \put(0,0){\includegraphics[width=\unitlength,page=1]{evictions-g5k_svg-tex.pdf}}%
    \put(0.06594867,0.04759085){\makebox(0,0)[t]{\lineheight{0}\smash{\begin{tabular}[t]{c}0\end{tabular}}}}%
    \put(0,0){\includegraphics[width=\unitlength,page=2]{evictions-g5k_svg-tex.pdf}}%
    \put(0.25219506,0.04759085){\makebox(0,0)[t]{\lineheight{0}\smash{\begin{tabular}[t]{c}10\end{tabular}}}}%
    \put(0,0){\includegraphics[width=\unitlength,page=3]{evictions-g5k_svg-tex.pdf}}%
    \put(0.43844148,0.04759085){\makebox(0,0)[t]{\lineheight{0}\smash{\begin{tabular}[t]{c}20\end{tabular}}}}%
    \put(0,0){\includegraphics[width=\unitlength,page=4]{evictions-g5k_svg-tex.pdf}}%
    \put(0.62468787,0.04759085){\makebox(0,0)[t]{\lineheight{0}\smash{\begin{tabular}[t]{c}30\end{tabular}}}}%
    \put(0,0){\includegraphics[width=\unitlength,page=5]{evictions-g5k_svg-tex.pdf}}%
    \put(0.81093424,0.04759085){\makebox(0,0)[t]{\lineheight{0}\smash{\begin{tabular}[t]{c}40\end{tabular}}}}%
    \put(0.50362769,0.01923651){\makebox(0,0)[t]{\lineheight{0}\smash{\begin{tabular}[t]{c}Evictions\end{tabular}}}}%
    \put(0,0){\includegraphics[width=\unitlength,page=6]{evictions-g5k_svg-tex.pdf}}%
    \put(0.02218076,0.29883138){\makebox(0,0)[lt]{\lineheight{0}\smash{\begin{tabular}[t]{l}Rescheduling Actions Taken\end{tabular}}}}%
    \put(0,0){\includegraphics[width=\unitlength,page=7]{evictions-g5k_svg-tex.pdf}}%
    \put(0.83824368,0.10082077){\makebox(0,0)[lt]{\lineheight{0}\smash{\begin{tabular}[t]{l}Buoyancy\end{tabular}}}}%
  \end{picture}%
\endgroup%

%% file: content/75.limitations.tex
\subsection{Threats to Validity}

Even though we have taken utmost care to ensure the validity of our results, there are some 
threats to their validity which we discuss here.

\subsubsection{Internal Threats}

The internal threats to our study stem from the limitations of our
available resources and the design of our experiments. We have taken care to ensure
that our experiments emulate a realistic cloud environment. However, we have not
been able to verify our approach in a real production cloud environment.

The selection of our workloads is designed to represent a wide range
of different types of cloud native applications with different
resource behavior. It is possible that other workloads may exhibit different behavior,
in ways that we have not captured in our experiments. 

Finally, our experiments are conducted on a set of nodes with a homogeneous hardware configuration. 
It is possible that other hardware configurations may exhibit different behavior,
for example due to variations in functionality between 
platforms and hardware generations~\cite{sohalCloserLookIntel2022}.

\subsubsection{External Threats}

Following the internal threats, we also identify some 
threats to the continued validity of our study beyond our control.
Our solution relies on hardware counters to compute resource and buoyancy scores, and
on the availability of cache allocation features to control the cache allocation of workloads.
These features vary between hardware platforms and generations, 
and may be removed or changed in future hardware. If no alternative features are available, 
our solution may not be applicable in future hardware.

Similarly, the cloud native nature of our implementation builds upon a number of
current software, such as Kubernetes, Prometheus, and Telegraf. These platforms
and technologies are under active development, and may change in ways that 
break our implementation. The foundational design of our solution is not 
dependent on any specific software, however, and can be adapted to future 
platforms and functionality.

%% file: content/80.conclusion.tex
\section{Conclusion}

In this paper, we have presented a holistic scheduling and resource management
solution that leverages resource congestion and application-level insights to
create a more efficient and balanced execution environment for compute clusters.
By leveraging the workload buoyancy framework, we supply resource owners the 
necessary insights to make informed decisions about resource allocation and scheduling
that balances the needs of both the applications and the underlying infrastructure.
Through extensive evaluation in a Kubernetes environment, we have demonstrated that our approach 
increases the percentage of workloads operating within their SLOs from $7.74\%$ to $87.8\%$
on average in a high congestion scenario compared to the baseline while
simultaneously reducing the size of violations when they occur. We show
that our approach reduces resource congestion and cluster fragmentation while
only taking a small amount of corrective rescheduling actions.

%% file: bib/references.bib
@misc{larssonWorkloadBuoyancyKeeping2026,
	title = {Workload Buoyancy: Keeping Apps Afloat by Identifying Shared Resource Bottlenecks},
	doi = {10.48550/arXiv.2602.22852},
	author = {Larsson, Oliver and Metsch, Thijs and Klein, Cristian and Elmroth, Erik},
	date = {2026-02},
	eprinttype = {arxiv},
	eprint = {2602.22852 [cs]},
}

@inproceedings{giannakopoulos_performance-aware_2025,
	title = {Performance-Aware Scheduling and Load Balancing for Edge/Cloud Systems},
	doi = {10.1109/IC2E65552.2025.00047},
	booktitle = {{IEEE} International Conference on Cloud Engineering ({IC}2E)},
	author = {Giannakopoulos, Panagiotis and van Knippenberg, Bart and Joshi, Kishor Chandra and Calabretta, Nicola and Exarchakos, George},
	date = {2025-09},
}

@article{da_silva_qos-driven_2020,
	title = {{QoS}-driven scheduling in the cloud},
	doi = {10.1186/s13174-020-00129-0},
	journaltitle = {Journal of Internet Services and Applications},
	author = {Da Silva, Giovanni Farias and Brasileiro, Francisco and Lopes, Raquel and Morais, Fabio and Carvalho, Marcus and Turull, Daniel},
	date = {2020},
}

@inproceedings{carvalho_qoe_2021,
	title = {{QoE}-Aware Container Scheduler for Co-located Cloud Environments},
	booktitle = {{IFIP}/{IEEE} International Symposium on Integrated Network Management ({IM})},
	author = {Carvalho, Marcos and Macedo, Daniel Fernandes},
	date = {2021-05},
}

@inproceedings{larsson_esther_2025,
	title = {{ESTHER}: Application-First Hardware-Level {QoS}-Enforcement for Cloud Native Environments},
	doi = {10.1109/CLOUD67622.2025.00018},
	booktitle = {{IEEE} 18th International Conference on Cloud Computing ({CLOUD})},
	author = {Larsson, Oliver and Metsch, Thijs and Klein, Cristian and Elmroth, Erik},
	date = {2025-07},
}

@inproceedings{zhao_rhythm_2020,
	title = {Rhythm: component-distinguishable workload deployment in datacenters},
	doi = {10.1145/3342195.3387534},
	booktitle = {Fifteenth European Conference on Computer Systems ({EuroSys})},
	author = {Zhao, Laiping and Yang, Yanan and Zhang, Kaixuan and Zhou, Xiaobo and Qiu, Tie and Li, Keqiu and Bao, Yungang},
	date = {2020-04},
}

@article{zhang_zeus_2021,
	title = {Zeus: Improving Resource Efficiency via Workload Colocation for Massive Kubernetes Clusters},
	doi = {10.1109/ACCESS.2021.3100082},
	journaltitle = {{IEEE} Access},
	author = {Zhang, Xiaolong and Li, Lanqing and Wang, Yuan and Chen, E. and Shou, Lidan},
	date = {2021-07},
}

@inproceedings{chen_olpart_2023,
	title = {{OLPart}: Online Learning based Resource Partitioning for Colocating Multiple Latency-Critical Jobs on Commodity Computers},
	doi = {10.1145/3552326.3567490},
	booktitle = {Eighteenth European Conference on Computer Systems ({EuroSys})},
	author = {Chen, Ruobing and Shi, Haosen and Li, Yusen and Liu, Xiaoguang and Wang, Gang},
	date = {2023-05},
}

@misc{byrne_survey_2018,
	title = {A Survey of Miss-Ratio Curve Construction Techniques},
	doi = {10.48550/arXiv.1804.01972},
	author = {Byrne, Daniel},
	date = {2018-04},
	eprinttype = {arxiv},
	eprint = {1804.01972 [cs]},
}

@inproceedings{park_copart_2019,
	title = {{CoPart}: Coordinated Partitioning of Last-Level Cache and Memory Bandwidth for Fairness-Aware Workload Consolidation on Commodity Servers},
	doi = {10.1145/3302424.3303963},
	booktitle = {Fourteenth European Conference on Computer Systems ({EuroSys})},
	author = {Park, Jinsu and Park, Seongbeom and Baek, Woongki},
	date = {2019-03},
}

@inproceedings{larsson_impact_2023,
	title = {The Impact of Directed Pod Eviction on Kubernetes Resource Utilization},
	doi = {10.1109/SOSE58276.2023.00016},
	booktitle = {{IEEE} International Conference on Service-Oriented System Engineering ({SOSE})},
	author = {Larsson, Oliver and Klein, Cristian and Elmroth, Erik},
	date = {2023-07},
}

@misc{larsson_beyond_2026,
	title = {Beyond the Limits: Flexible and Congestion-Aware Cluster Scheduling for the Cloud},
	doi = {10.48550/arXiv.2608.09308},
	number = {{arXiv}:2608.09308},
	publisher = {{arXiv}},
	author = {Larsson, Oliver and Metsch, Thijs and Klein, Cristian and Elmroth, Erik},
	date = {2026-08-10},
	eprinttype = {arxiv},
	eprint = {2608.09308 [cs.DC]},
}

@inproceedings{kastureTailbenchBenchmarkSuite2016,
	title = {Tailbench: a benchmark suite and evaluation methodology for latency-critical applications},
	doi = {10.1109/IISWC.2016.7581261},
	shorttitle = {Tailbench},
	booktitle = {{IEEE} International Symposium on Workload Characterization ({IISWC})},
	author = {Kasture, Harshad and Sanchez, Daniel},
	date = {2016-09},
}

@inproceedings{schroederOpenClosedCautionary2006,
	title = {Open Versus Closed: A Cautionary Tale},
	shorttitle = {Open Versus Closed},
	eventtitle = {Proceedings of the 3rd Conference on Networked Systems Design \& Implementation},
	author = {Schroeder, Bianca and Wierman, Adam and Harchol-Balter, Mor},
	date = {2006-05-01},
}

@software{inplacepodverticalscaling,
  author       = {Salloumis, Sotiris and Guo, Chunxia},
  title        = {{[FG:InPlacePodVerticalScalingExclusiveCPUs] Fix Static CPU management policy alongside InPlacePodVerticalScaling}},
  type         = {Pull Request \#129719},
  year         = {2026},
  url          = {https://github.com/kubernetes/kubernetes/pull/129719},
  urldate      = {2026-08-21}
}

@online{yuUserInterfaceResource,
	title = {User Interface for Resource Control feature (resctrl)},
	url = {https://docs.kernel.org/filesystems/resctrl.html},
	titleaddon = {The Linux Kernel documentation},
	author = {Yu, Fenghua and Luck, Tony and Shivappa, Vikas},
	urldate = {2026-08-21},
    date = {2026-05-08}
}

@article{larssonHardwareLevelQoSEnforcement2025,
	title = {Hardware-Level {QoS} Enforcement Features: Technologies, Use Cases, and Research Challenges},
	volume = {58},
	doi = {10.1145/3774317},
	pages = {145:1--145:35},
	number = {6},
	journaltitle = {{ACM} Comput. Surv.},
	author = {Larsson, Oliver and Metsch, Thijs and Klein, Cristian and Elmroth, Erik},
	date = {2025-12-17},
}

@inproceedings{sohalCloserLookIntel2022,
	title = {A Closer Look at Intel Resource Director Technology ({RDT})},
	doi = {10.1145/3534879.3534882},
	pages = {127--139},
	booktitle = {Proceedings of the 30th International Conference on Real-Time Networks and Systems},
	publisher = {{ACM}},
	author = {Sohal, Parul and Bechtel, Michael and Mancuso, Renato and Yun, Heechul and Krieger, Orran},
	date = {2022-06-07},
}

@inproceedings{freischuetzTUNATuningUnstable2025,
	title = {{TUNA}: Tuning Unstable and Noisy Cloud Applications},
	doi = {10.1145/3689031.3717480},
	booktitle = {Proceedings of the Twentieth European Conference on Computer Systems},
	publisher = {ACM},
	author = {Freischuetz, Johannes and Kanellis, Konstantinos and Kroth, Brian and Venkataraman, Shivaram},
	date = {2025-03-30},
}

@inproceedings{abediConductingRepeatableExperiments2017,
	title = {Conducting Repeatable Experiments in Highly Variable Cloud Computing Environments},
	doi = {10.1145/3030207.3030229},
	booktitle = {Proceedings of the 8th {ACM}/{SPEC} on International Conference on Performance Engineering},
	publisher = {ACM},
	author = {Abedi, Ali and Brecht, Tim},
	date = {2017-04-17},
}

@inproceedings{bashir_take_2021,
	title = {Take it to the limit: peak prediction-driven resource overcommitment in datacenters},
	doi = {10.1145/3447786.3456259},
	booktitle = {Proceedings of the {Sixteenth} {European} {Conference} on {Computer} {Systems}},
	publisher = {ACM},
	author = {Bashir, Noman and Deng, Nan and Rzadca, Krzysztof and Irwin, David and Kodak, Sree and Jnagal, Rohit},
	year = {2021},
}

@inproceedings{patel_clite_2020,
	title = {{CLITE}: {Efficient} and {QoS}-{Aware} {Co}-{Location} of {Multiple} {Latency}-{Critical} {Jobs} for {Warehouse} {Scale} {Computers}},
	doi = {10.1109/HPCA47549.2020.00025},
	booktitle = {{IEEE} {International} {Symposium} on {High} {Performance} {Computer} {Architecture} ({HPCA})},
	author = {Patel, Tirthak and Tiwari, Devesh},
	year = {2020},
}

@inproceedings{chen_parties_2019,
	title = {{PARTIES}: {QoS}-{Aware} {Resource} {Partitioning} for {Multiple} {Interactive} {Services}},
	doi = {10.1145/3297858.3304005},
	booktitle = {Proceedings of the {Twenty}-{Fourth} {International} {Conference} on {Architectural} {Support} for {Programming} {Languages} and {Operating} {Systems}},
	publisher = {ACM},
	author = {Chen, Shuang and Delimitrou, Christina and Martínez, José F.},
	year = {2019},
}

@inproceedings{aindalaAVOAEnergyResource2025,
	title = {{AVOA}: Energy and Resource Aware Scheduling Through Cloud Environment},
	doi = {10.1109/ICPCSN65854.2025.11035432},
	pages = {506--513},
	booktitle = {5th International Conference on Pervasive Computing and Social Networking ({ICPCSN})},
	author = {Aindala, Prashanthi and Shivakumar, K and Haripriya, Paladugula and Sivakumar, Ramagowni and Donthi, Rakesh Kumar and Medishetti, Santhosh Kumar},
	date = {2025-05},
}

@inproceedings{roySATORIEfficientFair2021,
	title = {{SATORI}: Efficient and Fair Resource Partitioning by Sacrificing Short-Term Benefits for Long-Term Gains},
	doi = {10.1109/ISCA52012.2021.00031},
	pages = {292--305},
	booktitle = {{ACM}/{IEEE} 48th Annual International Symposium on Computer Architecture ({ISCA})},
	author = {Roy, Rohan Basu and Patel, Tirthak and Tiwari, Devesh},
	date = {2021-06},
}

@article{chiuAnalysisIncreaseDecrease1989,
	title = {Analysis of the increase and decrease algorithms for congestion avoidance in computer networks},
	volume = {17},
	doi = {10.1016/0169-7552(89)90019-6},
	pages = {1--14},
	number = {1},
	journaltitle = {Computer Networks and {ISDN} Systems},
	author = {Chiu, Dah-Ming and Jain, Raj},
	date = {1989-06-10},
}
